\documentclass[1p,review,sort&compress]{elsarticle}
\usepackage{amsmath}
\usepackage{amssymb}
\usepackage[colorlinks=true,linkcolor=green,citecolor=red]{hyperref}
\usepackage{tikz}
\usetikzlibrary{mindmap}
\journal{Computer Physics Communications}

\begin{document}

\begin{frontmatter}

\title{$i$QIST: An open source continuous-time quantum Monte Carlo impurity solver toolkit}

\author[fribourg]{Li Huang\corref{cor1}} 
\ead{li.huang@unifr.ch}
\author[beijing]{Yilin Wang}
\author[beijing,toronto]{Zi Yang Meng}
\author[austin]{Liang Du}
\author[fribourg]{Philipp Werner}
\author[beijing]{Xi Dai}
\cortext[cor1]{Corresponding author}

\address[fribourg]{Department of Physics, University of Fribourg, 1700 Fribourg, Switzerland}
\address[beijing]{Beijing National Laboratory for Condensed Matter Physics, and Institute of Physics, Chinese Academy of Sciences, Beijing 100190, China}
\address[toronto]{Department of Physics, University of Toronto, Toronto, Ontario M5S 1A7, Canada}
\address[austin]{Department of Physics, The University of Texas at Austin, Austin, Texas 78712, USA}

\begin{abstract}
Quantum impurity solvers have a broad range of applications in theoretical studies of strongly correlated electron systems. Especially, they play a key role in dynamical mean-field theory calculations of correlated lattice models and realistic materials. Therefore, the development and implementation of efficient quantum impurity solvers is an important task. In this paper, we present an open source interacting quantum impurity solver toolkit (dubbed $i$QIST). This package contains several highly optimized quantum impurity solvers which are based on the hybridization expansion continuous-time quantum Monte Carlo algorithm, as well as some essential pre- and post-processing tools. We first introduce the basic principle of continuous-time quantum Monte Carlo algorithm and then discuss the implementation details and optimization strategies. The software framework, major features, and installation procedure for $i$QIST are also explained. Finally, several simple tutorials are presented in order to demonstrate the usage and power of $i$QIST.
\end{abstract}


\end{frontmatter}
\clearpage

\noindent\textbf{PROGRAM SUMMARY}

\noindent\textit{Program title:} $i$QIST

\noindent\textit{Catalogue identifier:} TO BE DONE

\noindent\textit{Program summary URL:} TO BE DONE

\noindent\textit{Program obtainable from:} CPC Program Library, Queen’s University, Belfast, N. Ireland

\noindent\textit{Licensing provisions:} GNU General Public Licence 3.0

\noindent\textit{No. of lines in distributed program, including test data, etc.:} 218579 lines

\noindent\textit{No. of bytes in distributed program, including test data, etc.:} 4613734.4 bytes

\noindent\textit{Distribution format:} tar.gz

\noindent\textit{Programming language:} Fortran 90 and Python

\noindent\textit{Computer:} Desktop PC, laptop, high performance computing cluster

\noindent\textit{Operating system:} Unix, Linux, Mac OS X, Windows

\noindent\textit{Has the code been vectorised or parallelized?:} Yes, it is parallelized by MPI and OpenMP

\noindent\textit{RAM:} Depends on the complexity of the problem

\noindent\textit{Classification:} 7.3

\noindent\textit{External routines/libraries used:} BLAS, LAPACK

\noindent\textit{Nature of problem:} Quantum impurity models were originally proposed to describe magnetic impurities in metallic hosts. In these models, the Coulomb interaction acts between electrons occupying the orbitals of the impurity atom. Electrons can hop between the impurity and the host, and in an action formulation, this hopping is described by a time-dependent hybridization function. Nowadays quantum impurity model have a broad range of applications, from the description of heavy fermion systems, and Kondo insulators, to quantum dots in nano-science. They also play an important role as auxiliary problems in dynamical mean-field theory and its diagrammatic extensions [1-3], where an interacting lattice model is mapped onto a quantum impurity model in a self-consistent manner. Thus, the accurate and efficient solution of quantum impurity models becomes an essential task.

\noindent\textit{Solution method:} The quantum impurity model can be solved by the numerically exact continuous-time quantum Monte Carlo method, which is the most efficient and powerful impurity solver for finite temperature simulations. In the $i$QIST software package, we implemented the hybridization expansion version of continuous-time quantum Monte Carlo algorithm. Both the segment representation and general matrix formalism are supported. The key idea of this algorithm is to expand the partition function diagrammatically in powers of the impurity-bath hybridization, and to stochastically sample these diagrams to all relevant orders using the Metropolis Monte Carlo algorithm.  For a detailed review of the continuous-time quantum Monte Carlo algorithms, please refer to [4].

\noindent\textit{Running time:} Depends on the complexity of the problem

\noindent\textit{References:}

[1] A. Georges, G. Kotliar, W. Krauth, and M. J. Rozenberg, Rev. Mod. Phys. 68, 13 (1996)

[2] G. Kotliar, S. Y. Savrasov, K. Haule, V. S. Oudovenko, O. Parcollet, and C. A. Marianetti, Rev. Mod. Phys. 78, 865 (2006)

[3] T. Maier, M. Jarrell, T. Pruschke, and M. H. Hettler, Rev. Mod. Phys. 77, 1027 (2005)

[4] E. Gull, A. J. Millis, A. I. Lichtenstein, A. N. Rubtsov, M. Troyer, and Philipp Werner, Rev. Mod. Phys. 83, 349 (2011)

\clearpage

\section{Introduction\label{sec:intro}}

In this paper we present $i$QIST (abbreviation for `interacting quantum impurity solver toolkit'), an open source project for recently developed hybridization expansion continuous-time quantum Monte Carlo impurity solvers~\cite{RevModPhys.83.349} and corresponding pre- and post-processing tools.

Dynamical mean-field theory (DMFT)~\cite{RevModPhys.68.13,RevModPhys.78.865} and its cluster extensions~\cite{RevModPhys.77.1027} play an important role in contemporary studies of correlated electron systems. The broad applications of this technique range from the study of Mott-Hubbard metal-insulator transitions~\cite{PhysRevB.54.8452}, unconventional superconductivity in Cu- and Fe-based superconductors~\cite{Yin2011,PhysRevLett.110.216405,PhysRevB.88.245110,Werner2012}, and non-Fermi liquid behaviors in multi-orbital systems~\cite{PhysRevLett.102.206407,PhysRevLett.108.216401,PhysRevLett.110.086401,PhysRevLett.101.166405}, to the investigation of anomalous transport properties of transition metal oxides~\cite{RevModPhys.70.1039}. For many of these applications, DMFT is the currently most powerful and reliable (sometimes the only) technique available and has in many cases produced new physical insights. Furthermore, the combination of \emph{ab initio} calculation methods (such as density function theory) with DMFT~\cite{RevModPhys.78.865} allows to capture the subtle electronic properties of realistic correlated materials, including those of partially filled $3d$- and $4d$-electron transition metal oxides, where lattice, spin and orbital degrees of freedom are coupled~\cite{RevModPhys.70.1039}.

The key idea of DMFT is to map the original correlated lattice model onto a quantum impurity model whose mean-field bath is determined self-consistently~\cite{RevModPhys.68.13,RevModPhys.77.1027,RevModPhys.78.865}. Thus, the central task of a DMFT simulation becomes the numerical solution of a quantum impurity problem. During the past several decades, many methods have been developed and tested as impurity solvers, including the exact diagonalization (ED)~\cite{PhysRevLett.72.1545}, equation of motion (EOM)~\cite{PhysRevB.50.7295}, Hubbard-I approximation (HIA)~\cite{PhysRevB.39.6962}, iterative perturbation theory (IPT)~\cite{PhysRevB.45.6479}, non-crossing approximation (NCA)~\cite{RevModPhys.59.845}, fluctuation-exchange approximation (FLEX)~\cite{PhysRevB.43.8044}, and quantum Monte Carlo (QMC)~\cite{PhysRevLett.56.2521,PhysRevLett.69.168}, etc. Among the methods listed above, the QMC method has several very important advantages, which makes it so far the most flexible and widely used impurity solver. First, it is based on the imaginary time action, in which the infinite bath has been integrated out. Second, it can treat arbitrary couplings, and can thus be applied to all kinds of phases including the metallic phase, insulating state, and phases with spontaneous symmetry breaking. Third, the QMC method is numerically exact with a ``controlled" numerical error. In other words, by increasing the computational effort the numerical error of the QMC simulation can be systematically reduced. For these reasons, the QMC algorithm is considered as the method of choice for many applications.

Several QMC impurity solvers have been developed in the past three decades. An important innovation was the Hirsch-Fye QMC (HF-QMC) impurity solver~\cite{PhysRevLett.56.2521,PhysRevLett.69.168}, in which the time axis is divided into small time steps and the interaction term in the Hamiltonian is decoupled on each time step by means of a discrete Hubbard-Stratonovich auxiliary field. HF-QMC has been widely used in the DMFT context~\cite{RevModPhys.68.13,RevModPhys.77.1027,RevModPhys.78.865}, but is limited by the discretization on the time axis and also by the form of the electronic interactions (usually only density-density interactions can be efficiently treated). Recently, a new class of more powerful and versatile QMC impurity solvers, continuous-time quantum Monte Carlo (CT-QMC) algorithms, have been invented~\cite{RevModPhys.83.349,PhysRevB.72.035122,Gull2008,PhysRevLett.97.076405,PhysRevB.74.155107,PhysRevB.75.155113}. In the CT-QMC impurity solvers, the partition function of the quantum impurity problem is diagrammatically expanded, and then the diagrammatic expansion series is evaluated by stochastic Monte Carlo sampling. The continuous-time nature of the algorithm means that operators can be placed at any arbitrary position on the imaginary time interval, so that time discretization errors can be completely avoided. Depending on how the diagrammatic expansion is performed, the CT-QMC approach can be further divided into interaction expansion (or weak coupling) CT-QMC (CT-INT)~\cite{PhysRevB.72.035122}, auxiliary field CT-QMC (CT-AUX)~\cite{Gull2008}, and hybridization expansion (or strong coupling) CT-QMC (CT-HYB)~\cite{PhysRevLett.97.076405,PhysRevB.74.155107,PhysRevB.75.155113}.

At present, CT-HYB is the most popular and powerful impurity solver, since it can be used to solve multi-orbital impurity models with general interactions at low temperature~\cite{RevModPhys.83.349}. In single-site DMFT calculations, the computational efficiency of CT-HYB is much higher than that of CT-INT, CT-AUX, and HF-QMC, especially when the interactions are intermediate or strong. However, in order to solve more complicated quantum impurity models (for example, five-band or seven-band impurity model with general interactions and spin-orbital coupling) efficiently, further improvements of the CT-HYB impurity solvers are needed. In recent years many tricks and optimizations have been explored and implemented to increase the efficiency and accuracy of the original CT-HYB algorithm, such as the truncation approximation~\cite{PhysRevB.75.155113}, Krylov subspace iteration~\cite{PhysRevB.80.235117}, orthogonal polynomial representation~\cite{PhysRevB.84.075145,PhysRevB.85.205106,PhysRevB.89.235128}, PS quantum number~\cite{PhysRevB.86.155158}, lazy trace evaluation~\cite{PhysRevB.90.075149}, skip-list technique~\cite{PhysRevB.90.075149}, matrix product state implementation~\cite{hir2014}, and sliding window sampling scheme~\cite{hir2014}, etc. As the state-of-the-art CT-HYB impurity solvers become more and more sophisticated and specialized, it is not easy anymore to master all their facets and build one's implementations from scratch. Hence, we believe that it is a good time to provide a CT-HYB software package for the DMFT community such that researchers can focus more on the physics problems, instead of spending much time on (re-)implementing in-house codes. In fact, there are some valuable efforts in this direction, such as TRIQS~\cite{triqs}, ALPS~\cite{hafermann,alps}, W2DYNAMICS~\cite{PhysRevB.86.155158}, DMFT\_W2K~\cite{PhysRevB.75.155113,PhysRevB.81.195107}, etc. The present implementation of the CT-HYB impurity solvers is a useful complement to the existing codes. The open source $i$QIST software package contains several well-implemented and thoroughly tested modern CT-HYB impurity solvers, and the corresponding pre- and post-processing tools. We hope the release of $i$QIST can promote the quick development of this research field.

The rest of this paper is organized as follows: In Sec.~\ref{sec:theory}, the basic theory of quantum impurity models, CT-QMC algorithms, and its hybridization expansion version are briefly introduced. The measurements of several important physical observables are presented. In Sec.~\ref{sec:implement}, the implementation details of $i$QIST are discussed. Most of the optimization tricks and strategies implemented in $i$QIST, including dynamical truncation, lazy trace evaluation, sparse matrix technique, PS quantum number, and subspace algorithms, etc., are reviewed. These methods ensure the high efficiency of $i$QIST. In Sec.~\ref{sec:overview}, we first present an overview on the software architecture and component framework. Then the main features of the $i$QIST software package, including the CT-HYB impurity solvers, the atomic eigenvalue solver, and the other auxiliary tools are presented. The compiling and installation procedures, and the basic usage of $i$QIST are introduced in Sec.~\ref{sec:install}. Section~\ref{sec:examples} shows several simple applications of $i$QIST, ranging from self-consistent single-site DMFT calculation to one-shot post-processing calculation. These examples serve as introductory tutorials. Finally, a short summary is given in Sec.~\ref{sec:conclusion} and the future development plans for the $i$QIST project are outlined as well.

\section{Basic theory and methods\label{sec:theory}}
In this section, we will present the basic principles of CT-QMC impurity solvers, with an emphasis on the hybridization expansion technique. For detailed derivations and explanations, please refer to Ref.~\cite{RevModPhys.83.349}. 

\subsection{Quantum impurity model\label{subsec:aim}}
The multi-orbital Anderson impurity model (AIM) can be written as $H_{\text{imp}} = H_{\text{loc}} + H_{\text{bath}} + H_{\text{hyb}}$, where
\begin{subequations}
\label{eq:aimp}
\begin{align}
& H_{\text{loc}} = \sum_{\alpha\beta} E_{\alpha\beta} d_{\alpha}^{\dagger} d_{\beta}+\sum_{\alpha\beta\gamma\delta} U_{\alpha\beta\gamma\delta} 
    d^{\dagger}_{\alpha}d^{\dagger}_{\beta} d_{\gamma} d_{\delta}, \\
& H_{\text{hyb} } = \sum_{\textbf{k}\alpha\beta} V^{\alpha\beta}_{\textbf{k}} c_{\textbf{k}\alpha}^{\dagger} d_{\beta} + h.c., \\
& H_{\text{bath}} = \sum_{\textbf{k}\alpha} \epsilon_{\textbf{k}\alpha} c_{\textbf{k}\alpha}^{\dagger} c_{\textbf{k}\alpha}.
\end{align}
\end{subequations}
In Eq.~(\ref{eq:aimp}), Greek letters in the subscripts denote a combined spin-orbital index, the operator $d_\alpha^{\dagger}$ ($d_\alpha$) is creating (annihilating) an electron with index $\alpha$ on the impurity site, while $c_{\textbf{k}\alpha}^{\dagger}$ ($c_{\textbf{k}\alpha}$) is the creation (annihilation) operator for conduction band (bath) electron with spin-orbital index $\alpha$ and momentum $\textbf{k}$. The first term in $H_{\text{loc}}$ is the general form of the impurity single particle term with impurity level splitting and inter-orbital hybridization. This term can be built by crystal field (CF) splitting and/or spin-orbit coupling (SOC), etc. The second term in $H_{\text{loc}}$ is the Coulomb interaction term which can be parameterized by intra(inter)-band Coulomb interactions $U$ $(U')$ and Hund's rule coupling $J$ or Slater integral parameters $F^{k}$. The hybridization term $H_{\text{hyb}}$ describes the process of electrons hopping from the impurity site to the environment and back. $H_{\text{bath}}$ describes the non-interacting bath. This Anderson impurity model is solved self-consistently in the DMFT calculations~\cite{RevModPhys.68.13,RevModPhys.78.865}.

\subsection{Principles of continuous-time quantum Monte Carlo algorithm\label{subsec:ctqmc}}
We first split the full Hamiltonian $H_{\text{imp}}$ into two separate parts, $H_{\text{imp}} = H_1 + H_2$, then treat $H_2$ as a perturbation term, and expand the partition function $\mathcal{Z} = \text{Tr} e^{-\beta H}$ in powers of $H_2$,
\begin{equation}
\label{eq:partition}
\mathcal{Z} = \sum_{n=0}^{\infty} \int_{0}^{\beta} \cdots \int_{\tau_{n-1}}^\beta \omega(\mathcal{C}_n),
\end{equation}
with
\begin{equation}
\label{eq:weight}
\omega(\mathcal{C}_n)=d\tau_1 \cdots d\tau_n \text{Tr}\left\{ e^{-\beta H_1}[-H_2(\tau_n)]\cdots [-H_2(\tau_1)]\right\},
\end{equation}
where $H_2(\tau)$ is defined in the interaction picture with $H_2(\tau) = e^{\tau H_1} H_2 e^{-\tau H_1}$. Each term in Eq.~(\ref{eq:partition}) can be regarded as a diagram or configuration (labelled by $\mathcal{C}$), and $\omega(\mathcal{C}_n)$ is the diagrammatic weight of a specific order-$n$ configuration. Next we use a stochastic Monte Carlo algorithm to sample the terms of this series. In the CT-INT and CT-AUX impurity solvers~\cite{PhysRevB.72.035122,Gull2008}, the interaction term is the perturbation term, namely, $H_2 = H_{\text{int}}$, while $H_2 = H_{\text{hyb}}$ is chosen for the CT-HYB impurity solver~\cite{PhysRevLett.97.076405}. In the intermediate and strong interaction region, CT-HYB is much more efficient than CT-INT and CT-AUX. This is also the main reason that we only implemented the CT-HYB impurity solvers in the $i$QIST software package.

\subsection{Hybridization expansion\label{subsec:cthyb}}
In the hybridization expansion algorithm, due to fact that $H_1$ does not mix the impurity and bath states, 
the trace in Eq.~(\ref{eq:weight}) can be written as $\text{Tr} = \text{Tr}_d \text{Tr}_c$. 
As a result, we can split the weight of each configuration as 
\begin{equation}
\omega(\mathcal{C}_n) = \omega_{d}(\mathcal{C}_n) \omega_{c}(\mathcal{C}_n) \prod\limits_{i=1}^{n} d\tau_i.
\end{equation}
$\omega_{d}(\mathcal{C}_n)$ is the trace over impurity operators ($\text{Tr}_d$), $\omega_{c}(\mathcal{C}_n)$ is the trace over bath operators ($\text{Tr}_c$). Further, since Wick's theorem is applicable for the $\omega_c(\mathcal{C}_n)$ part, we can represent it as a determinant of a matrix $\mathcal{Z}_{\text{bath}}\mathcal{M}^{-1}$ with $\mathcal{Z}_{\text{bath}}=\text{Tr}_c e^{-\beta H_{\text{bath}}}$ and $(\mathcal{M}^{-1})_{ij} = \Delta(\tau_i - \tau_j)$. The $\omega_{d}(\mathcal{C}_n)$ part can be expressed using the segment representation when $[n_{\alpha}, H_{\text{loc}}] = 0$~\cite{PhysRevLett.97.076405}. However, if this condition is not fulfilled, we have to calculate the trace explicitly, which is called the general matrix algorithm~\cite{PhysRevB.75.155113,PhysRevB.74.155107}. The explicit calculation of the trace for a large multi-orbital AIM with general interactions is computationally expensive. Many tricks and strategies have been implemented in the $i$QIST software package to address this challenge. Please refer to Sec.~\ref{sec:implement} for more details.

In this package, we used importance sampling and the Metropolis algorithm to evaluate Eq.~(\ref{eq:partition}). 
The following four local update procedures, with which the ergodicity of Monte Carlo algorithm is guaranteed, are used to generate the Markov chain: 
\begin{itemize}
\item Insert a pair of creation and annihilation operators in the time interval $[0,\beta)$.
\item Remove a pair of creation and annihilation operators from the current configuration.
\item Select a creation operator randomly and shift its position in the time interval $[0,\beta)$.
\item Select an annihilation operator randomly and shift its position in the time interval $[0,\beta)$.
\end{itemize}
In the Monte Carlo simulations, sometimes the system can be trapped in some (for example symmetry-broken) state. In order to avoid unphysical trapping, we also consider the following two global updates:
\begin{itemize}
\item Swap the operators of randomly selected spin up and spin down flavors.
\item Swap the creation and annihilation operators globally.
\end{itemize}

\subsection{Physical observables\label{subsec:obs}}
Many physical observables are measured in our CT-HYB impurity solvers. Here we provide a list of them.

\underline{Single-particle Green's function $G(\tau)$}

The most important observable is the single-particle Green's function $G(\tau)$, which is measured using the elements of the matrix $\mathcal{M}$, 
\begin{align}
\label{eq:gt}
G(\tau) = \left\langle \frac{1}{\beta} \sum_{ij}\delta^{-}(\tau, \tau_i - \tau_j) \mathcal{M}_{ji}\right\rangle,
\end{align}
with 
\begin{align}
\delta^{-}(\tau, \tau') = 
\begin{cases} 
\delta(\tau - \tau'), & \tau' > 0, \\
-\delta(\tau - \tau' + \beta), & \tau' < 0.
\end{cases}
\end{align}
Note that in the $i$QIST software package, the low-frequency Matsubara Green's function $G(i\omega_n)$ is also measured directly, instead of being calculated from $G(\tau)$ using Fourier transformation.

\underline{Two-particle correlation function $\chi_{\alpha\beta}(\tau_a,\tau_b,\tau_c,\tau_d)$}

The two-particle correlation functions are often used to construct lattice susceptibilities within DMFT and diagrammatic extensions of DMFT. However, the measurements of two-particle correlation functions are a nontrivial task~\cite{PhysRevB.83.085102} as it is very time-consuming to obtain good quality data, and most of the previous publications in this field are restricted to measurements of two-particle correlation functions in one-band models. Thanks to the development of efficient CT-HYB algorithms, the calculation of two-particle correlation functions for multi-orbital impurity models now becomes affordable~\cite{PhysRevB.84.075145,PhysRevB.89.235128,PhysRevB.85.205106}. In the $i$QIST software package, we implemented the measurement for the two-particle correlation function $\chi_{\alpha\beta}(\tau_a,\tau_b,\tau_c,\tau_d)$, which is defined as follows:
\begin{equation}
\chi_{\alpha\beta}(\tau_a,\tau_b,\tau_c,\tau_d)
= \langle c_{\alpha}(\tau_a)c^{\dagger}_{\alpha}(\tau_b)c_{\beta}(\tau_c)c^{\dagger}_{\beta}(\tau_d)\rangle.
\end{equation}
Due to memory restrictions, the actual measurement is performed in frequency space, for which we use the following definition of the Fourier transform:
\begin{align}
\chi_{\alpha\beta}(\omega,\omega',\nu) &= \frac{1}{\beta}
\int^{\beta}_{0}d\tau_a\int^{\beta}_{0}d\tau_b\int^{\beta}_{0}d\tau_c\int^{\beta}_{0}d\tau_d
\nonumber\\
&\times \chi_{\alpha\beta}(\tau_a,\tau_b,\tau_c,\tau_d) 
e^{i(\omega+\nu)\tau_a}e^{-i\omega\tau_b}e^{-i\omega'\tau_c}e^{-i(\omega'+\nu)\tau_d}.
\end{align}
where $\omega$ and $\omega'$ [$\equiv (2n+1)\pi\beta$] are fermionic frequencies, and $\nu$ is bosonic ($\equiv 2n\pi/\beta$).

\underline{Local irreducible vertex functions $\Gamma_{\alpha\beta}(\omega,\omega',\nu)$}

From the two-particle correlation function $\chi_{\alpha\beta}(\omega,\omega',\nu)$, the local irreducible vertex function $\Gamma_{\alpha\beta}(\omega,\omega',\nu)$ can be calculated easily, via the Bethe-Salpeter equation~\cite{PhysRevB.86.125114,PhysRevB.85.205106,PhysRevB.89.235128}:
\begin{equation}
\Gamma_{\alpha\beta}(\omega,\omega',\nu) = 
\frac{\chi_{\alpha\beta}(\omega,\omega',\nu) 
- \beta[G_\alpha(\omega+\nu)G_\beta(\omega')\delta_{\nu,0} 
- G_\alpha(\omega+\nu) G_\beta(\omega') \delta_{\alpha\beta}\delta_{\omega\omega'}]}
{G_\alpha(\omega+\nu)G_\alpha(\omega)G_\beta(\omega')G_\beta(\omega'+\nu)}.
\end{equation}
The $G(i\omega_n)$ and $\Gamma_{\alpha\beta}(\omega,\omega',\nu)$ are essential inputs for the diagrammatic extensions of DMFT, such as the dual fermions (DF)~\cite{PhysRevB.77.033101} and dynamical vertex approximation (D$\Gamma$A) \cite{PhysRevB.75.045118} codes.

\underline{Impurity self-energy function $\Sigma(i\omega_n)$}

The self-energy $\Sigma(i\omega_n)$ is calculated using Dyson's equation 
\begin{equation}
\Sigma(i\omega_n) = G^{-1}_{0}(i\omega_n) - G^{-1}(i\omega_n),
\end{equation}
 or measured using the so-called improved estimator~\cite{PhysRevB.89.235128,PhysRevB.85.205106}. Note that in the current implementation the latter approach only works when the segment representation is used.

\underline{Histogram of the perturbation expansion order}

We record the histogram of the perturbation expansion order $k$, which can be used to evaluate the kinetic energy via Eq.~(\ref{eq:kin}) below.

\underline{Occupation number and double occupation number}

The orbital occupation number $\langle n_\alpha\rangle$ and double occupation number $\langle n_\alpha n_\beta \rangle$ are measured. From them we can calculate for example the charge fluctuation $\sqrt{\langle N^2 \rangle - \langle N \rangle^2}$, where $N$ is the total occupation number:
\begin{equation}
N = \sum_{\alpha} n_{\alpha}.
\end{equation}

\underline{Spin-spin correlation function}

For a system with spin rotational symmetry, the expression for the spin-spin correlation function reads 
\begin{equation}
\chi_{ss}(\tau) = \langle S_{z}(\tau) S_{z}(0) \rangle,
\end{equation}
where $S_{z} = n_{\uparrow} - n_{\downarrow}$. From it we can calculate the effective magnetic moment:
\begin{equation}
\mu_{\text{eff}} = \int^{\beta}_{0}d\tau \chi_{ss}(\tau).
\end{equation}

\underline{Orbital-orbital correlation function}

The expression for the orbital-orbital correlation function reads 
\begin{equation}
\chi^{nn}_{\alpha\beta}(\tau) = \langle n_{\alpha}(\tau) n_{\beta}(0) \rangle.
\end{equation}

\underline{Kinetic energy}

In DMFT, the expression for the kinetic energy of the lattice model reads
\begin{equation}
\label{eq:kin}
E_\text{kin} = -\frac{1}{\beta} \langle k \rangle,
\end{equation}
where $k$ is the perturbation expansion order.

\underline{Atomic state probability}

The expression for the atomic state probability is 
\begin{equation}
p_{\Gamma} = \langle |\Gamma \rangle \langle \Gamma| \rangle,
\end{equation}
where $\Gamma$ is the atomic state.

\section{Implementations and optimizations\label{sec:implement}}

In this section, we will focus on the implementation details and discuss the optimization tricks adopted in the $i$QIST software package.

\subsection{Development platform\label{subsec:platform}}
The major part of the $i$QIST software package was developed with the modern Fortran 90 language. We extensively used advanced language features in the Fortran 2003/2008 standard such as an object oriented programming style (polymorphic, inheritance, and module, etc.) to improve the readability and re-usability of the source codes. The compiler is fixed to the Intel Fortran compiler. We can not guarantee that the $i$QIST can be compiled successfully with other Fortran compilers. Some auxiliary scripts, pre- and post-processing tools are written using the Python language and Bash shell scripts. These scripts and tools act like a glue. They are very flexible and can be easily extended or adapted to deal with various problems. These Python codes can run properly under the Python 2.x or 3.x runtime environment.

Since $i$QIST is a big software development project, we use Git as the version control system, and the source codes are hosted in a remote server. The developers pull the source codes from the server into their local machines, and then try to improve them. Once the modification is completed, the source codes can be pushed back to the server and merged with the master branch. Then the other developers can access them and use them immediately to start further developments. The members of our developer team can access the code repository anywhere and anytime.

\subsection{Orthogonal polynomial representation\label{subsec:legendre}}
Boehnke \emph{et al.}~\cite{PhysRevB.84.075145} proposed to use Legendre polynomials to improve the measurements of single-particle and two-particle Green's functions. Thanks to the Legendre polynomial representation, the numerical noise and memory space needed to store the Green's function are greatly reduced.

The imaginary time Green's function $G(\tau)$ is expressed using the Legendre polynomial $P_n(x)$ defined in [-1,1]:
\begin{equation}
\label{eq:gt_new}
G(\tau) = \frac{1}{\beta} \sum_{n \leq 0} \sqrt{2n + 1} P_{n}[x(\tau)] G_n,
\end{equation}
where $n$ is the order of Legendre polynomial, $G_{n}$ is the expansion coefficient, $x(\tau)$
maps $\tau \in [0,\beta]$ to $x \in [-1,1]$:
\begin{equation}
x(\tau) = \frac{2\tau}{\beta} - 1.
\end{equation}
Using the orthogonality relations of Legendre polynomials, we obtain 
\begin{equation}
\label{eq:gn_old}
G_n = \sqrt{2n + 1} \int^{\beta}_{0} d\tau P_{n}[x(\tau)] G(\tau).
\end{equation}
If we substitute Eq.~(\ref{eq:gt}) into Eq.~(\ref{eq:gn_old}), we get 
\begin{equation}
G_n = -\frac{\sqrt{2n + 1}}{\beta} \left\langle \sum^{k}_{i=1} \sum^{k}_{j=1}
\mathcal{M}_{ji} \tilde{P}_{n}(\tau^e_i - \tau^s_j) \right\rangle,
\end{equation}
where
\begin{equation}
\tilde{P}_n(\tau) = 
\begin{cases}
P_n [x(\tau)], & \tau > 0, \\
-P_n [x(\tau + \beta)], & \tau < 0, \\
\end{cases}
\end{equation}
and $\tau^{s}$ and $\tau^{e}$ denote the positions of creation and annihilation operators on the imaginary time axis, respectively. We can also express the Matsubara Green's function $G(i\omega_n)$ using Legendre polynomials:
\begin{equation}
\label{eq:gw_new}
G(i\omega_m) = \sum_{n \leq 0} T_{mn} G_n.
\end{equation}
The transformation matrix $T_{mn}$ is defined as
\begin{equation}
T_{mn} = (-1)^m i^{n+1} \sqrt{2n + 1} j_n \left[\frac{(2m + 1)\pi}{2}\right],
\end{equation}
where $j_n(z)$ is the spheric Bessel function. Actually, in the Monte Carlo simulation, only the expansion coefficients $G_n$ are measured. When the calculation is finished, the final Green's function can be evaluated using Eq.~(\ref{eq:gt_new}) and Eq.~(\ref{eq:gw_new}). It is worthwhile to note that the $T_{mn}$ do not depend on the inverse temperature $\beta$, so that we can calculate and store the matrix elements beforehand to save computer time.

It is easy to extend this formalism to other orthogonal polynomials. For example, in the $i$QIST software package, we not only implemented the Legendre polynomial representation, but also the Chebyshev polynomial representation. In the Chebyshev polynomial representation, the imaginary time Green's function $G(\tau)$ is expanded as follows:
\begin{equation}
G(\tau) = \frac{2}{\beta} \sum_{n \leq 0} U_n [x({\tau})]G_{n},
\end{equation}
where the $U_n(x)$ denote the second kind of Chebyshev polynomials and $x \in [-1,1]$. The equation for the expansion coefficients $G_n$ is:
\begin{equation}
G_n = -\frac{2}{\pi\beta} \left\langle  \sum^{k}_{i=1} \sum^{k}_{j=1} 
\mathcal{M}_{ji} 
\tilde{U}_{n}(\tau^e_i - \tau^s_j)
\sqrt{1 - \tilde{x}(\tau^e_i - \tau^s_j)^2}
\right\rangle,
\end{equation}
where
\begin{equation}
\tilde{U}_n (x) = 
\begin{cases}
U_n[x(\tau)], & \tau > 0, \\
-U_n[x(\tau+\beta)], & \tau < 0, \\
\end{cases}
\end{equation}
and
\begin{equation}
\tilde{x}(\tau) = 
\begin{cases}
x(\tau), & \tau > 0, \\
x(\tau + \beta), & \tau < 0. \\
\end{cases}
\end{equation}
Unfortunately, there is no explicit expression for $G(i\omega_n)$ [like Eq.~(\ref{eq:gw_new})] in the Chebyshev polynomial representation.

\subsection{Improved estimator for the self-energy function and vertex function\label{subsec:self}}
Recently, Hafermann \emph{et al.} proposed efficient measurement procedures for the self-energy and vertex functions within the CT-HYB algorithm~\cite{PhysRevB.85.205106,PhysRevB.89.235128}. In their method, some higher-order correlation functions (related to the quantities being sought through the equation of motion) are measured. For the case of density-density interactions, the segment algorithm is available~\cite{PhysRevLett.97.076405}. Thus, the additional correlators can be obtained essentially without additional computational cost. When the calculations are completed, the required self-energy function and vertex function can be evaluated analytically. 

The improved estimator for the self-energy function can be expressed in the following form:
\begin{equation}
\label{eq:self-energy}
\Sigma_{ab}(i\omega_n) = \frac{1}{2} 
\sum_{ij} G^{-1}_{ai}(i\omega_n) (U_{jb} + U_{bj}) F^{j}_{ib}(i\omega_n),
\end{equation}
where $U_{ab}$ is the Coulomb interaction matrix element. The expression for the new two-particle correlator $F^{j}_{ab}(\tau - \tau')$ reads
\begin{equation}
F^{j}_{ab}(\tau-\tau') 
= -\langle \mathcal{T} d_{a}(\tau) d^{\dagger}_{b}(\tau') n_{j}(\tau') \rangle,
\end{equation}
and $F^{j}_{ab}(i\omega_n)$ is its Fourier transform. The actual measurement formula is
\begin{equation}
\label{eq:fj}
F^{j}_{ab}(\tau - \tau') = 
-\frac{1}{\beta}
\left\langle
\sum_{\alpha\beta = 1}^{k} 
\mathcal{M}_{\beta\alpha}\delta^{-}(\tau-\tau', \tau^{e}_{\alpha} - \tau^{s}_{\beta})
n_{j}(\tau^s_\beta)\delta_{a,\alpha}\delta_{b,\beta}
\right\rangle.
\end{equation}
The measurement formula for the vertex function can be found in the original paper~\cite{PhysRevB.85.205106,PhysRevB.89.235128}. Note that when the Coulomb interaction is frequency-dependent, Eq.~(\ref{eq:self-energy}) and (\ref{eq:fj}) should be modified slightly~\cite{PhysRevB.89.235128}. As one can see, this equation for $F^{j}_{ab}(\tau - \tau')$ looks quite similar to Eq.~(\ref{eq:gt}). Thus we use the same method to measure $F^{j}_{ab}(\tau - \tau')$ and finally get the self-energy function via Eq.~(\ref{eq:self-energy}). Here, the matrix element $n_{j}(\tau^s_\beta)$ (one or zero) denotes whether or not the flavor $j$ is occupied (whether or not a segment is present) at time $\tau^s_\beta$.

This method can be combined with the orthogonal polynomial representation~\cite{PhysRevB.84.075145} as introduced in the previous subsection to suppress fluctuations and filter out the Monte Carlo noise. Using this technique, we can obtain the self-energy and vertex functions with unprecedented accuracy, which leads to an enhanced stability in the analytical continuations of those quantities~\cite{PhysRevB.85.205106}.
 
\subsection{Subspaces and symmetry\label{subsec:subspace}}
As mentioned before, for a Hamiltonian $H_{\text{loc}}$ with general interactions the evaluation of local trace is heavily time-consuming,
\begin{equation}
\label{equ:tr1}
\omega_{d}(\mathcal{C}) = 
\text{Tr}_{\text{loc}} (T_{2k+1}F_{2k}T_{2k} \cdots F_{1}T_{1}),
\end{equation}  
where $T=e^{-\tau H_{\text{loc}}}$ is time evolution operator, $F$ is fermionic creation or annihilation operator, and $k$ is expansion order for the current diagrammatic configuration $\mathcal{C}$. The straightforward method to evaluate this trace is to insert the complete eigenstates $\{ \Gamma \}$ of $H_{\text{loc}}$ into the RHS of Eq.~(\ref{equ:tr1}), then 
\begin{equation}\label{equ:tr2}
\omega_{d}(\mathcal{C}) = \sum_{\{\Gamma_{1} \cdots \Gamma_{2k}\}} 
            \langle\Gamma_{1}|T_{2k+1}|\Gamma_{1}\rangle
            \langle\Gamma_{1}|F_{2k}|\Gamma_{2k}\rangle \langle\Gamma_{2k}|T_{2k}|\Gamma_{2k}\rangle \cdots  
            \langle\Gamma_{2}|F_{1}|\Gamma_{1}\rangle
            \langle\Gamma_{1}|T_{1}|\Gamma_{1}\rangle.
\end{equation}
Thus, we must do $4k+1$ matrix-matrix multiplications with the dimension of the Hilbert space of $H_{\text{loc}}$. This method is robust but very slow for large multi-orbital impurity model as the dimension of the matrix is impractically large for 5- and 7-band systems, and the expansion order $k$ is large as well.

Actually, the matrices of the fermion operators ($F$-matrix) are very sparse due to the symmetry of $H_{\text{loc}}$. We can take advantage of this to speed up the matrix-matrix multiplications. We exploit the symmetry of $H_{\text{loc}}$ to find some good quantum numbers (GQNs) and divide the full Hilbert space of $H_{\text{loc}}$ with very large dimension into much smaller subspaces labeled by these GQNs~\cite{RevModPhys.83.349}. We call such a subspace $|\alpha\rangle$ a superstate~\cite{PhysRevB.75.155113} which consists of all the $n_{\alpha}$ eigenstates of this subspace, $|\alpha\rangle=\{ \Gamma_{1}, \Gamma_{2}, \cdots, \Gamma_{n_{\alpha}}\}$. The $F$-matrix element can only be nonzero between pairs of superstates with different values of GQNs. One fermion operator may bring one initial superstate $|\alpha\rangle$ to some other final superstates $|\beta\rangle$,
\begin{equation}\label{equ:next_sect}
F|\alpha\rangle= |\beta\rangle,
\end{equation}
or outside of the full Hilbert space. We have to carefully choose the GQNs to make sure that for a fixed initial superstate $|\alpha\rangle$ and a fixed fermion operator, there is one and only one final superstate $|\beta\rangle$ if it doesn't go outside of the full Hilbert space. Given an arbitrary diagrammatic configuration, starting with a superstate $|\alpha_{1}\rangle$, there will be only one possible evolution path. That is,
\begin{equation}
|\alpha_{1}\rangle \xrightarrow{F_{1}} 
|\alpha_{2}\rangle \xrightarrow{F_{2}} 
|\alpha_{3}\rangle \xrightarrow{F_{3}} 
|\alpha_{4}\rangle \cdots 
|\alpha_{2k-1}\rangle \xrightarrow{F_{2k-1}} 
|\alpha_{2k}\rangle   \xrightarrow{F_{2k}} 
|\alpha_{1}\rangle.
\end{equation}
The path may break at some point because it goes outside of the full Hilbert space or violates the Pauli principle. For a successful path starting with $|\alpha_{1}\rangle$, its contribution to the local trace is
\begin{eqnarray}
\text{Tr}_{\alpha_{1}} = 
\sum_{\{ \Gamma_{\alpha_{1}} \cdots \Gamma_{\alpha_{2k}}\}}
&&\langle\Gamma_{\alpha_{1}}|T_{2k+1}|\Gamma_{\alpha_{1}}\rangle
\langle\Gamma_{\alpha_{1}}|F_{2k}|\Gamma_{\alpha_{2k}}\rangle
\langle\Gamma_{\alpha_{2k}}|T_{2k}|\Gamma_{\alpha_{2k}}\rangle \cdots \nonumber \\
&&\langle\Gamma_{\alpha_{2}}|F_{1}|\Gamma_{\alpha_{1}}\rangle
\langle\Gamma_{\alpha_{1}}|T_{1}|\Gamma_{\alpha_{1}}\rangle,
\end{eqnarray}
where $\{ \Gamma_{\alpha_{i}} \}$ are the eigenstates of subspace $\alpha_{i}$. Thus, the final local trace should be
\begin{equation}
\omega_{d}(\mathcal{C}) = \sum_{i} \text{Tr}_{\alpha_{i}}.
\end{equation} 
As a result, the original $4k+1$ matrix-matrix multiplications with large dimension reduces to several times $4k+1$ matrix-matrix multiplications with much smaller dimensions, resulting in a huge speedup.

\begin{table}[t]
\caption{The GQNs supports for various types of local Hamiltonians $H_{\text{loc}}$. \label{table:good}}
\centering
\begin{tabular}{cccc}
\hline
\hline
 GQNs                   & Kanamori-$U$ & Slater-$U$ &  SOC \\
\hline
$N, S_{z}$              & Yes          & Yes        & No   \\ 
$N, S_{z}$, PS          & Yes          & No         & No   \\ 
$N, J_{z}$              & Yes          & Yes        & Yes  \\
$N$                     & Yes          & Yes        & Yes  \\ 
\hline
\hline
\end{tabular}
\end{table}

In our codes, we implemented several GQNs schemes for different types of local Hamiltonians $H_{\text{loc}}$, as summarized in Table~\ref{table:good}. For $H_{\text{loc}}$ without SOC, we have two choices: (1) with Slater parameterized Coulomb interaction matrix, we use the total occupation number $N$, the $z$ component of total spin $S_{z}$ as GQNs; (2) with Kanamori parameterized Coulomb interaction matrix, besides $N$ and $S_{z}$, we can use another powerful GQN, the so-called PS number~\cite{PhysRevB.86.155158}. It is defined as,
\begin{equation}
\text{PS} = \sum_{\alpha=1}^{N_{\text{orb}}} \\
             (n_{\alpha\uparrow}-n_{\alpha\downarrow})^2 \times 2^{\alpha},
\end{equation}
where $\alpha$ is the orbital index, $\{\uparrow, \downarrow\}$ is spin index, $n_{\alpha\uparrow}$ and $n_{\alpha\downarrow}$ are the orbital occupancy numbers. The PS number labels the occupation number basis with the same singly occupied orbitals. With its help, the dimensions of the subspaces become very small, such that we can treat 5-band Kanamori parameterized interaction systems efficiently without any approximations. For $H_{\text{loc}}$ with SOC, we can use the total occupancy number $N$ and the $z$ component of total angular momentum $J_{z}$ as GQNs. We summarize the total number of subspaces, maximum and mean dimensions of subspaces for different GQNs schemes and multi-orbital impurity models in Table.~\ref{table:dim}. Obviously, using these GQNs can largely reduce the dimension of the $F$-matrix, and make accurate DMFT calculations for complex electronic systems (such as the $d$- and $f$-electron materials) possible. 

\begin{table}[t]
\caption{The total number of subspaces $N$, maximum and mean dimensions of subspaces for different GQNs schemes and multi-orbital models. \label{table:dim}} 
\centering
\begin{tabular}{ccccc}
\hline
\hline
                & 2-band       & 3-band       & 5-band       & 7-band          \\
\hline
GQNs            & $N$/max/mean & $N$/max/mean & $N$/max/mean & $N$/max/mean    \\ 
\hline
$N, S_{z}$      &  9/4/1.78    & 16/9/4.00    & 36/100/28.44 & 64/1225/256.00  \\
$N, S_{z}$, PS  &  14/2/1.14   & 44/3/1.45    & 352/10/2.91  & 2368/35/6.92    \\
$N, J_{z}$      &  -           & 26/5/2.46    & 96/37/10.67  & 246/327/66.60   \\
$N$             &  5/6/3.20    & 7/20/9.14    & 11/252/93.09 & 15/3432/1092.27 \\ 
\hline
\hline
\end{tabular}
\end{table}

\subsection{Truncation approximation\label{subsec:truncation}}
As discussed in Sec.~\ref{subsec:subspace}, although we have used GQNs to split the full Hilbert space with very large dimension into blocks with smaller dimensions [for cases such as 7-band systems with GQNs ($N$, $J_{z}$) and 5-band systems with GQN ($N$)], the dimensions of some blocks are still too large and the number of blocks is too high, so that it is still very expensive to evaluate the local trace. K. Haule proposed in Ref.~\cite{PhysRevB.75.155113} to discard some high-energy states because they are rarely visited. For example, for 7-band system with only 1 electron (like Ce metal), only states with occupancy $N=0$, 1, 2 will be frequently visited, and states with occupancy $N>2$ can be truncated completely to reduce the large Hilbert space to a very small one. Of course, this truncation approximation may cause some bias because a frequently visited state may be accessed via an infrequently visited state. Therefore, one should be cautious when adopting the truncation approximation, and for example run some convergence tests. 

Currently, we adopted two truncation schemes in our codes. The first scheme relies on the cut-off of the occupation number. We just keep those states whose occupation numbers are close to the nominal valence and skip the other states, as shown in the above Ce metal example. This scheme is quite robust if the charge fluctuations are small enough, such as in the case of a Mott insulating phase. Another scheme is to dynamically truncate the states with very low probability based on statistics which is recorded during the Monte Carlo sampling. This scheme is not very stable, so one needs to use it with caution.

\subsection{Lazy trace evaluation\label{subsec:lazy}}
The diagrammatic Monte Carlo sampling algorithm consists of the following steps: (1) Propose an update for the current diagrammatic configuration. (2) Calculate the acceptance probability $p$ according to the Metropolis-Hasting algorithm,
\begin{equation}
p = \text{min} \left(1, \frac{A^\prime}{A} \left| \frac{\omega_{c}}{\omega_{c}^{\prime}}\right| 
     \left|\frac{\omega_{d}}{\omega_{d}^{\prime}} \right|\right),
\end{equation}
where, $A$ is the proposal probability for the current update and $A^\prime$ for the inverse update, $\omega_{c}$ and $\omega_{c}^{\prime}$ are the determinants for the new and old configurations, respectively, and $\omega_{d}$ and $\omega_{d}^{\prime}$ are the local traces for the new and old configurations, respectively. (3) Generate a random number $r$. If $p>r$, the proposed update is accepted, otherwise it is rejected. (4) Update the current diagrammatic configuration if the proposed update is accepted. It turns out that for CT-HYB, $p$ is usually low ($1\% \sim 20\%$), especially in the low temperature region. On the other hand, the calculation of $p$ involves a costly local trace evaluation. To avoid wasting computation time when the acceptance probability is very low, in the subspace algorithm, we implemented the so-called lazy trace evaluation proposed in Ref.~\cite{PhysRevB.90.075149}.

The basic idea of the lazy trace evaluation is simple. For the proposed Monte Carlo move, we first generate a random number $r$. Then, instead of calculating the local trace from scratch to determine $p$, we calculate bounds for $\left|\text{Tr}_{\text{loc}}\right|$,
\begin{equation}
\left|\omega_{d}\right| = \left|\text{Tr}_{\text{loc}}\right| \leq \sum_{i} \left|\text{Tr}_{i}\right| \leq \sum_{i} B_{i},
\end{equation}
where $B_i \geq \left|\text{Tr}_{i}\right|$. $B_{i}$ is a product of some chosen matrix norms of $T$ and $F$ matrices: 
\begin{equation}
B_i = C  \left\| T_{2k+1} 
\right\| \left\| F_{2k} \right\| \left\| T_{2k} \right\| \cdots   
\left\| F_{1} \right\| \left\| T_{1} \right\| \geq
\left|\text{Tr}(T_{2k+1}F_{2k} T_{2k}\cdots F_{1}T_{1})\right|,
\end{equation}
where $C$ is a parameter depending on the specific type of matrix norm, and $\left\| \cdot \right\|$ denotes a matrix norm.
If $\text{Tr}_{i^\prime}$ denotes the exact traces of some subspaces, then we have 
\begin{equation}
\left| \left|\text{Tr}_{\text{loc}}\right| - \sum_{i^\prime}\left|\text{Tr}_{i^\prime}\right| \right| 
\leq \sum_{i \neq i^\prime} B_{i}.
\end{equation}
Thus, we can determine the upper $p_{\text{max}}$ and lower $p_{\text{min}}$ bounds of $p$ as
\begin{equation}
\begin{aligned}
p_{\text{max}}=R \left(\sum_{i^\prime}\left|\text{Tr}_{i^\prime}\right| + \sum_{i \neq i^\prime} B_{i}\right),\\
p_{\text{min}}=R \left(\sum_{i^\prime}\left|\text{Tr}_{i^\prime}\right| - \sum_{i \neq i^\prime} B_{i}\right),
\end{aligned}
\end{equation} 
where $R=\frac{A^\prime}{A} \left| \frac{\omega_{c}}{\omega_{c}^{\prime}}\right| \left|\frac{1}{\omega_{d}^{\prime}} \right|$.
If $r>p_{\text{max}}$, we reject this move immediately. If $r<p_{\text{min}}$, we accept the move and calculate the determinant and local trace from scratch. If $ p_{\text{min}} < r < p_{\text{max}} $, we refine the bounds by calculating the local trace of one more subspace $\text{Tr}_{i}$ until we can reject or accept the move. The calculation of these bounds involves only simple linear algebra calculations of matrix norms which cost little computation time, and one refining operation involves only one subspace trace evaluation. On average, it saves a lot of computation time, as confirmed by our benchmarks.

\subsection{Divide-and-conquer and sparse matrix tricks\label{subsec:sparse}}
\begin{figure}[tp]
\centering
\includegraphics[width=1.0\textwidth]{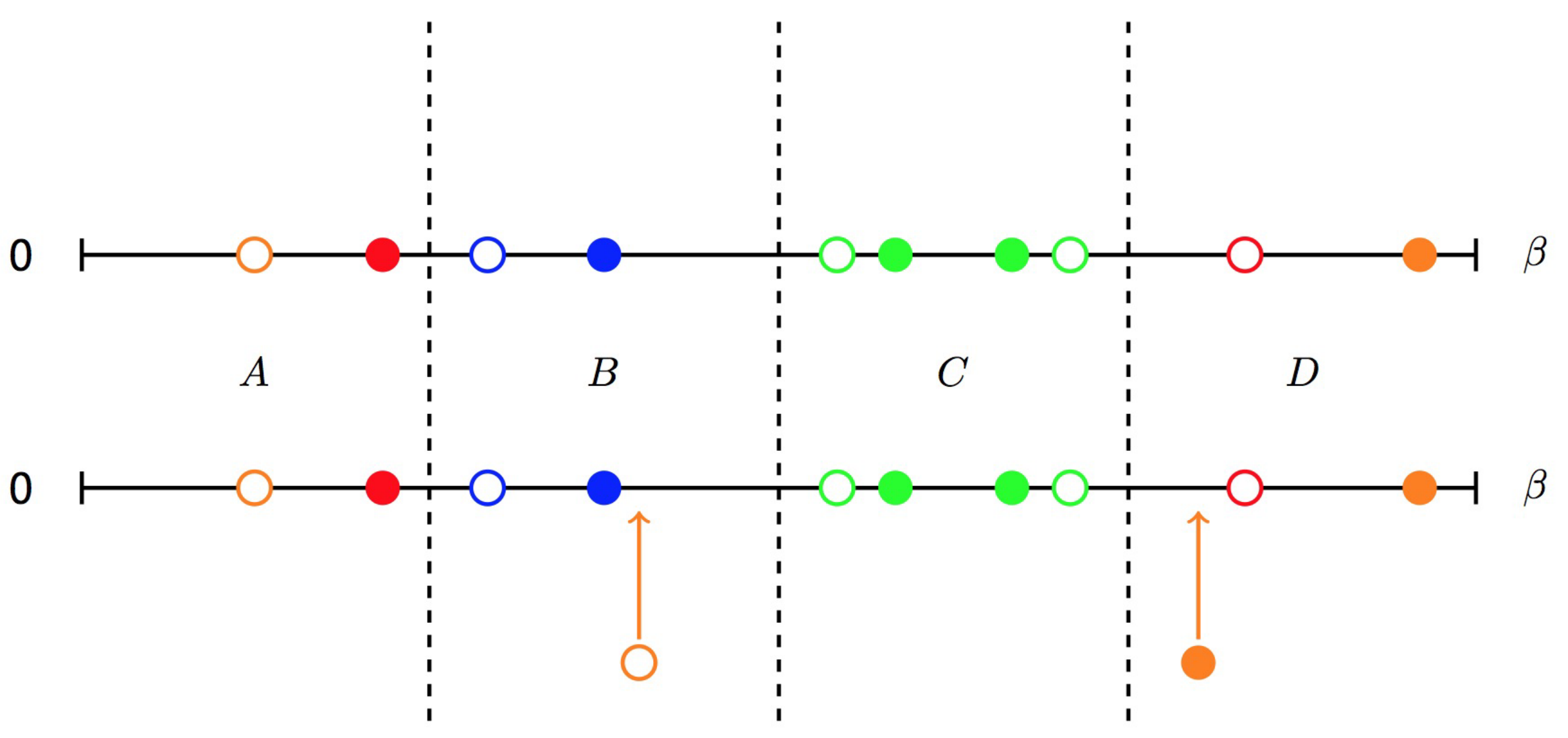}
\caption{Illustration of the divide-and-conquer algorithm. The imaginary time interval $[0,\beta)$ is split into four parts with equal length by vertical dashed lines. The open (filled) circles mean creation (annihilation) operators. The color is used to distinguish different flavors. It shows that a creation operator is inserted into the $B$ part, while a annihilation operator is inserted into the $D$ part. \label{fig:div}}
\end{figure}

The Monte Carlo updates, such as inserting (removing) a pair of creation and annihilation operators, usually modify the diagrammatic configuration locally. Based on this fact, we implemented a divide-and-conquer algorithm to speed up the trace evaluation. As illustrated in Fig.~\ref{fig:div}, we divide the imaginary time interval $[0,\beta)$ into a few parts with equal length. For each part, there will be zero or nonzero fermion operators, and we save their matrix products when evaluating the local trace in the beginning. In the next Monte Carlo sampling, we first determine which parts may be modified or influenced, and then for these parts we recalculate the matrix products from scratch and save them again. For the unchanged parts, we will leave them alone. Finally, we will multiply the contributions of all parts to obtain the final local trace. By using this divide-and-conquer trick, we can avoid redundant computations and speed up the calculation of the acceptance probability $p$. This trick can be combined with the GQNs algorithm and lazy trace evaluation to achieve a further speedup. 

If direct matrix-matrix multiplications are used when evaluating the local trace, the $F$-matrix must be very sparse. Thus, we can convert them into sparse matrices in compressed sparse row (CSR) format, and then the sparse matrix multiplication can be applied to obtain a significant speedup.

\subsection{Random number generators\label{subsec:rng}}
Fast, reliable, and long period pseudo-random number generators are a key factor for Monte Carlo simulations. Currently, the most popular random number generator is the Mersenne Twister which was developed by Matsumoto and Nishimura~\cite{Matsumoto:1998}. Its name derives from the fact that its period length is chosen to be a Mersenne prime. In the $i$QIST software package, we implemented the commonly used version of Mersenne Twister, MT19937. It has a very long period of $2^{19937}-1$.

The Mersenne Twister is a bit slow by today's standards. So in 2006, a variant of Mersenne Twister, the SIMD-oriented Fast Mersenne Twister (SFMT) was introduced~\cite{sfmt:2008}. It was designed to be fast when it runs on 128-bit SIMD. It is almost twice as fast as the original Mersenne Twister and has better statistics properties. We also implemented it in the $i$QIST software package, and use it as the default random number generator.

\subsection{Parallelization\label{subsec:mpi}}
All of the CT-HYB impurity solvers in the $i$QIST software package are parallelized by MPI. The strategy is very simple. In the beginning, we launch $n$ processes simultaneously. The master process is responsible for reading input data and configuration parameters, and broadcasts them among the child processes. And then each child process will perform Monte Carlo samplings and measure physical observables independently. After all the processes finish their jobs, the master process will collect the measured quantities from all the processes and average them to obtain the final results. Apart from that, no additional inter-process communication is needed. Thus, we can anticipate that the parallel efficiency will be very good, and near linear speedups are possible, as long as the number of thermalization steps is small compared to the total number of Monte Carlo steps. In practical calculations, we usually fix the number of Monte Carlo steps $N_{\text{sweep}}$ done by each process, and launch as many processes as possible. Given that the number of processes is $N_{\text{proc}}$, then the total number of Monte Carlo samplings should be $N_{\text{proc}}N_{\text{sweep}}$. Naturally, the more processes we use, the more accurate data we can obtain.

For some specific tasks, such as the measurement of two-particle quantities, fine-grained parallelism is necessary. Thus, we further parallelized them with the OpenMP multi-thread technology. So, in order to attain ideal speedup, we have to carefully choose suitable numbers of MPI processes and OpenMP threads.

\section{Features\label{sec:overview}}
In this section, we will introduce the software architecture and component framework of $i$QIST. The major features of its components are presented in detail.

\subsection{Software architecture\label{subsec:framework}}
\begin{figure}[tp]
\centering
\includegraphics[width=1.0\textwidth]{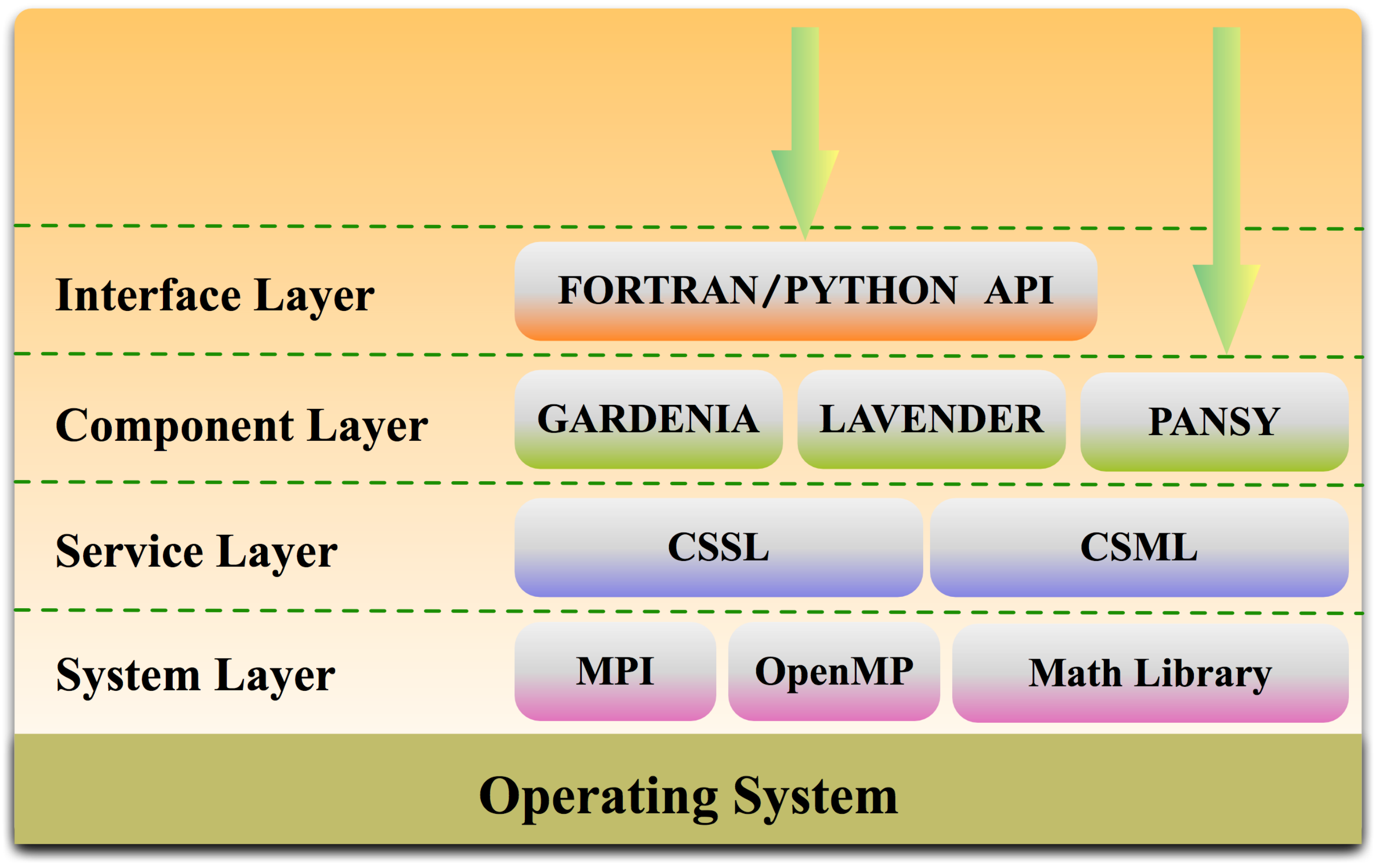}
\caption{The hierarchical structure of the $i$QIST software package. Note that in the component layer, not all of the components are listed due to space limitations. See the main text for detailed explanations. \label{fig:framework}}
\end{figure}

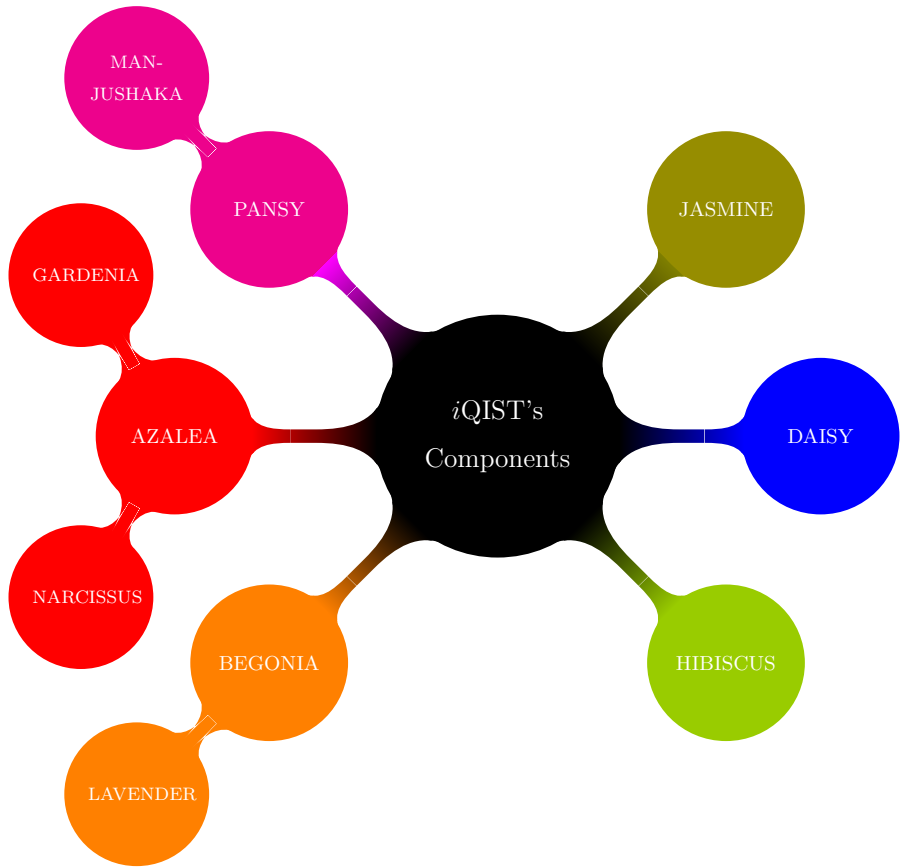
\begin{figure}[ht]
\centering
\scalebox{0.85}{\begin{tikzpicture}[mindmap, concept color=black, text=white]
    \node[concept, minimum size=2.6cm] {$i$QIST's\\ Components}
        child[concept color=blue, grow=000, minimum size=2.4cm] {node[concept]{DAISY}}
        child[concept color=olive, grow=045, minimum size=2.4cm] {node[concept]{JASMINE}}
        child[concept color=lime!80!black, grow=315, minimum size=2.4cm] {node[concept]{HIBISCUS}}
        child[concept color=red, grow=180, minimum size=2.4cm] {node[concept]{AZALEA}
            child[concept, grow=120, minimum size=2.2cm] {node[concept]{GARDENIA}}
            child[concept, grow=240, minimum size=2.2cm] {node[concept]{NARCISSUS}}}
        child[concept color=orange, grow=225, minimum size=2.4cm] {node[concept]{BEGONIA}
            child[concept, grow=225, minimum size=2.2cm] {node[concept]{LAVENDER}}}
        child[concept color=magenta, grow=135,minimum size=2.4cm] {node[concept]{PANSY}
            child[concept, grow=135, minimum size=2.2cm] {node[concept]{MAN\-JUSHAKA}}};
\end{tikzpicture}}
\caption{Schematic picture for the $i$QIST's components. Components on the LHS are the CT-HYB solvers, \texttt{JASMINE} is the atomic eigenvalue solver, \texttt{DAISY} is a HF-QMC solver, and \texttt{HIBISCUS} contains the other pre- and post-processing tools. \label{fig:componentlayer}}
\end{figure}

To solve a quantum impurity model is not a straightforward job. Besides the necessary quantum impurity solvers, we need some auxiliary programs or tools. The $i$QIST is an all-in-one software package, which can be used to solve a broad range of quantum impurity problems. It is a collection of various codes and scripts whose core components contain about 120000 lines of code. 

The software architecture of $i$QIST is slightly involved. In Fig.~\ref{fig:framework}, we use a layer model to illustrate it. The bottom layer is the operating system (OS). In principle, the $i$QIST is OS-independent. It can run properly on top of Unix/Linux, Mac OS X, FreeBSD, and Windows. The second layer is the system layer, which contains highly optimized linear algebra math libraries (such as BLAS and LAPACK) and parallelism supports (such as MPI and OpenMP). The third layer is the service layer. In this layer, we implemented some commonly used modules and subroutines. They are called common service subroutine library (CSSL) and common service module library (CSML), respectively. They provide a useful interface between the system layer and the component layer and facilitate the development of core components. The features of CSSL and CSML include basic data structures (stack and linked list), random number generators, sparse matrix manipulations, linear algebra operations, string processing, linear interpolation, numerical integration, fast Fourier transformation (FFT), etc. 

The core part of $i$QIST is in the fourth layer -- the component layer -- which contains various impurity solvers and auxiliary tools as shown in Fig.~\ref{fig:componentlayer}. At present, $i$QIST contains ten different components. They are \texttt{AZALEA}, \texttt{GARDENIA}, \texttt{NARCISSUS}, \texttt{BEGONIA}, \texttt{LAVENDER}, \texttt{PANSY}, \texttt{MANJUSHAKA}, \texttt{DAISY}, \texttt{JASMINE}, and \texttt{HIBISCUS}. Here, \texttt{AZALEA}, \texttt{GARDENIA}, \texttt{NARCISSUS}, \texttt{BEG\-O\-NIA}, \texttt{LAVENDER}, \texttt{PANSY}, and \texttt{MANJUSHAKA} are all CT-HYB impurity solver components (as shown in the LHS of Fig.~\ref{fig:componentlayer}), and \texttt{DAISY} is a HF-QMC impurity solver component. \texttt{JASMINE} is an atomic eigenvalue solver. \texttt{HIBISCUS} is a collection of several pre- and post-processing tools, including the maximum entropy method, stochastic analytical continuation, Pad\'{e} approximation, and Kramers-Kronig transformation, etc. For more details about these components, please consult the following sections. 

The top layer is the interface layer or user layer. On the one hand, we can execute $i$QIST's components directly as usual. On the other hand, we can also invoke $i$QIST's components from other languages. The role of $i$QIST's components becomes a library or subroutine. To achieve this goal, in the interface layer, we offer the Fortran/Python language bindings for most of the $i$QIST components, so that we can develop our own codes on top of $i$QIST and consider it as a computational engine in black box.

\subsection{CT-HYB impurity solvers\label{subsec:azalea}}
\begin{table}[ht]
\centering
\caption{The models supported by various CT-HYB impurity solvers in the $i$QIST software package. In this and the following tables, the CT-HYB impurity solvers are abbreviated using the first capital letter of their names. For example, \texttt{A} denotes the \texttt{AZALEA} component. \label{tab:feature_model}}
\begin{tabular}{lr}
\hline
\hline
Models & CT-HYB \\
\hline
Density-density interaction & \texttt{A, G, N, B, L, P, M}\\
General Coulomb interaction (Slater or Kanamori schemes) & \texttt{B, L, P, M} \\
Spin-orbit coupling interaction & \texttt{B, L, P, M} \\
Crystal field splitting & \texttt{A, G, N, B, L, P, M} \\
Hubbard-Holstein model & \texttt{N} \\
Frequency-dependent (retarded) interaction & \texttt{N} \\
\hline
\hline
\end{tabular}
\end{table}

\begin{table}[ht]
\centering
\caption{The measurement tricks used by various CT-HYB impurity solvers in the $i$QIST software package. \label{tab:feature_tricks}}
\begin{tabular}{lr}
\hline
\hline
Measurement tricks & CT-HYB \\
\hline
Orthogonal polynomial representation (Legendre and Chebyshev types) & \texttt{G, N, L, M} \\
Improved estimator for self-energy and vertex functions & \texttt{G, N} \\
\hline
\hline
\end{tabular}
\end{table}

\begin{table}[ht]
\centering
\caption{The trace evaluation algorithms supported by various CT-HYB impurity solvers in the $i$QIST software package. \label{tab:feature_fast}}
\begin{tabular}{lr}
\hline
\hline
Trace algorithms & CT-HYB \\
\hline
Segment representation algorithm & \texttt{A, G, N} \\
Divide-and-conquer algorithm & \texttt{B, L, P, M} \\
Sparse matrix multiplication & \texttt{B, L} \\
Good quantum numbers & \texttt{P, M} \\
Skip-lists trick & \texttt{M} \\
Lazy trace evaluation & \texttt{M} \\
Dynamical truncation approximation & \texttt{M} \\
\hline
\hline
\end{tabular}
\end{table}

\begin{table}[ht]
\centering
\caption{The observables measured by various CT-HYB impurity solvers in the $i$QIST software package. \label{tab:feature_observables}}
\begin{tabular}{lr}
\hline
\hline
Physical observables & CT-HYB \\
\hline
Single-particle Green's function $G(\tau)$ & \texttt{A, G, N, B, L, P, M} \\
Single-particle Green's function $G(i\omega_n)$ & \texttt{A, G, N, B, L, P, M} \\
Two-particle correlation function $\chi(\omega, \omega', \nu)$ & \texttt{G, N, L, M} \\
Local irreducible vertex function $\Gamma(\omega, \omega', \nu)$ & \texttt{G, N, L, M} \\
Pair susceptibility $\Gamma_{\text{pp}}(\omega, \omega', \nu)$ & \texttt{G, N, L, M} \\
Self-energy function $\Sigma(i\omega_n)$ & \texttt{A, G, N, B, L, P, M} \\
Histogram of perturbation expansion order & \texttt{A, G, N, B, L, P, M} \\
Kinetic and potential energies & \texttt{A, G, N, B, L, P, M} \\
(Double) occupation numbers, magnetic moment & \texttt{A, G, N, B, L, P, M} \\
Atomic state probability & \texttt{A, G, N, B, L, P, M} \\
Spin-spin correlation function & \texttt{G, N} \\
Orbital-orbital correlation function & \texttt{G, N} \\
Autocorrelation function and autocorrelation time & \texttt{G, N, L, M} \\
\hline\hline
\end{tabular}
\end{table}
As mentioned before, the $i$QIST software package contains seven CT-HYB impurity solvers (as schematically shown in Fig.~\ref{fig:componentlayer}). In this subsection, in order to help the users to choose a suitable CT-HYB impurity solver, we briefly discuss their main features, pros, and cons. The main results are also summarized in Tab.~\ref{tab:feature_model}-\ref{tab:feature_observables} for a quick query.

When the Coulomb interaction term in the local Hamiltonian $H_{\text{loc}}$ is of density-density type, $H_{\text{loc}}$ becomes a diagonal matrix in the occupation number basis. In this case, the CT-HYB impurity solver is extremely efficient if the so-called segment picture (or segment representation)~\cite{PhysRevLett.97.076405,RevModPhys.83.349} is adopted. Thus, we implemented the segment algorithm in the \texttt{AZALEA}, \texttt{GARDENIA}, and \texttt{NARCISSUS} components.

In the \texttt{AZALEA} component, we only implemented the basic segment algorithm and very limited physical observables are measured. It is the simplest and the most efficient code. In fact, it is the development prototype of the other CT-HYB components, and usually used to test some experimental features. In the \texttt{GARDENIA} component, we add more features on the basis of the \texttt{AZALEA} component. For example, we can use the orthogonal polynomial technique to improve the numerical accuracy and suppress stochastic noise in the Green's function~\cite{PhysRevB.84.075145}. The self-energy function can be measured with the improved estimator method~\cite{PhysRevB.89.235128,PhysRevB.85.205106}. More single-particle and two-particle correlation functions are measured. Though \texttt{GARDENIA} is much more powerful than \texttt{AZALEA}, it is a bit less efficient. The features of the \texttt{NARCISSUS} component are almost the same as those of the \texttt{GARDENIA} component. In addition, it can be used to deal with dynamically screened interactions~\cite{PhysRevLett.104.146401,Werner2012}. In other words, the Coulomb interaction $U$ need not to be a static value any more, but can be frequency-dependent. Thus, it is used for example in extended-DMFT calculations~\cite{PhysRevB.87.125149}. Note that since the Hubbard-Holstein model can be mapped in DMFT onto a dynamical Anderson impurity model~\cite{PhysRevLett.99.146404}, it can be solved using the \texttt{NARCISSUS} component as well.

When the local Hamiltonian $H_{\text{loc}}$ contains general Coulomb interaction terms, there is no simple expression for the $\omega_d(\mathcal{C}_n)$ and the segment representation is not applicable any more. At that time, the general matrix formulation~\cite{PhysRevB.74.155107,PhysRevB.75.155113}, which is implemented in the \texttt{BEGONIA}, \texttt{LAVENDER}, \texttt{PANSY}, and \texttt{MANJUSHAKA} components, should be used. Each of these components has its own features and targets specific systems.

In the \texttt{BEGONIA} component, we implemented the direct matrix-matrix multiplications algorithm. We adopted the divide-and-conquer scheme and sparse matrix technique to speed up the calculation. This component can be used to deal with impurity models with up to $3$ bands with fairly good efficiency. However, it is not suitable for 5- and 7-band systems. In the \texttt{LAVENDER} component, we implemented all the same algorithms as in the \texttt{BEGONIA} component. Besides, we implemented the orthogonal polynomial representation to improve the measurement quality of physical quantities. Some two-particle quantities are also measured. This component should also only be used to conduct calculations for $1 \sim 3$ bands systems. But it can produce measurements of very high quality with small additional cost. In the \texttt{PANSY} component, we exploited the symmetries of $H_{\text{loc}}$ and applied the GQNs trick to accelerate the evaluation of local trace. This algorithm is general and doesn't depend on any details of the GQNs, so it can support all the GQNs schemes which fulfill the conditions discussed in Sec.~\ref{subsec:subspace}. We also adopted the divide-and-conquer algorithm to speed it up further. This component can be used to study various impurity models ranging from 1-band to 5-band with fairly good efficiency. However, it is still not suitable for 7-band models. In the \texttt{MANJUSHAKA} component, we implemented all the same algorithms as the \texttt{PANSY} component. Besides, we implemented the lazy trace evaluation~\cite{PhysRevB.90.075149} to speed up the Monte Carlo sampling process. It can gain quite high efficiency, and is extremely useful in the low temperature region. We also implemented a smart algorithm to truncate some high-energy states dynamically in the Hilbert space of $H_{\text{loc}}$ to speed up the trace evaluation further. This algorithm is very important and efficient (in many situations it is necessary) for dealing with 7-band systems. We implemented the orthogonal polynomial representation to improve the measurements of key observables as well. By using all of these tricks, the computational efficiency of the \texttt{MANJUSHAKA} component for multi-orbital impurity models with general Coulomb interaction is very high. We believe that it can be used to study most quantum impurity systems ranging from 1-band to 7-band.

\subsection{Atomic eigenvalue solver\label{subsec:jasmine}}
When the Coulomb interaction is general in the local Hamiltonian $H_{\text{loc}}$, as discussed above, we have to diagonalize $H_{\text{loc}}$ in advance to obtain its eigenvalues, eigenvectors, and the $F$-matrix. In general, the local Hamiltonian is defined as
\begin{equation}
H_{\text{loc}} = H_{\text{int}} + H_{\text{cf}} + H_{\text{soc}},
\end{equation}
where $H_{\text{int}}$ means the Coulomb interaction term, $H_{\text{cf}}$ the CF splitting term, and $H_{\text{soc}}$ the SOC interaction. The \texttt{JASMINE} component is used to solve this Hamiltonian and generate necessary inputs for some CT-HYB impurity solvers (i.e., \texttt{BEGONIA}, \texttt{LAVENDER}, \texttt{PANSY}, and \texttt{MANJUSHAKA} components).

The \texttt{JASMINE} component will build $H_{\text{loc}}$ in the Fock representation at first. For the Coulomb interaction term $H_{\text{int}}$, both Kanamori parameterized and Slater parameterized forms are supported. In other words, we can use $U$ and $J$, or Slater integrals $F^{k}$ to define the Coulomb interaction matrix as we wish. For the CF splitting term $H_{\text{cf}}$, both diagonal and non-diagonal elements are accepted. The SOC term $H_{\text{soc}}$ is defined as follows,
\begin{equation}
H_{\text{soc}} = \lambda \sum_i \vec{\mathbf{l}}_i \cdot \vec{\mathbf{s}}_i,
\end{equation}
where $\lambda$ is the strength of the SOC. Note that the SOC term can only be activated for the 3-, 5-, and 7-band systems.

Next, the \texttt{JASMINE} component will diagonalize $H_{\text{loc}}$ to get all eigenvalues and eigenvectors. There are two running modes for \texttt{JASMINE}. (1) It diagonalizes $H_{\text{loc}}$ in the full Hilbert space directly to obtain the eigenvalues $E_{\alpha}$ and eigenvectors $\Gamma_{\alpha}$, then the $F$-matrix is built from the eigenvectors,
\begin{equation}
(F_{i})_{\alpha,\beta} = \langle\Gamma_{\alpha}|F_{i}|\Gamma_{\beta}\rangle,
\end{equation}
where $i$ is the flavor index. The eigenvalues and $F$-matrix will be fed into the \texttt{BEGONIA} and \texttt{LAVENDER} components as necessary input data. (2) It diagonalizes each subspace of $H_{\text{loc}}$ according to the selected GQNs. Currently, four GQNs schemes for various types of $H_{\text{loc}}$ are supported, which are summarized in Table~\ref{table:good}. \texttt{JASMINE} also builds indices to record the evolution sequence depicted in Eq.~(\ref{equ:next_sect}). According to the indices, it builds the $F$-matrix between two different subspaces. The eigenvalues, the indices, and the $F$-matrix will be collected and written into an external file (atom.cix), which will be read by the \texttt{PANSY} and \texttt{MANJUSHAKA} components.

Apart from this, the \texttt{JASMINE} component will also generate the matrix elements of some physical operators, such as $\vec{L}^2$, $L_{z}$, $\vec{S}^2$, $S_{z}$, $\vec{J}^2$, and $J_{z}$, etc. They can be used by the other post-processing codes to analyze the averaged expectation value of these operators.

\subsection{Auxiliary tools\label{subsec:hibiscus}}
In the \texttt{HIBISCUS} component, many auxiliary tools are provided to deal with the output data of the CT-HYB impurity solvers. Here we briefly describe some of these tools:

\underline{Maximum entropy method}

In the Monte Carlo community, the maximum entropy method~\cite{mem:1996} is often used to extract the spectral function $A(\omega)$ from the imaginary time Green's function $G(\tau)$. Thus, in the \texttt{HIBISCUS} component, we implemented the standard maximum entropy algorithm. In the Extended-DMFT calculations, sometimes we have to perform an analytical continuation for the retarded interaction function $\mathcal{U}(i\nu)$ to obtain $\mathcal{U}(\nu)$~\cite{PhysRevB.90.195114}. So we developed a modified version of the maximum entropy method to enable this calculation.

\underline{Stochastic analytical continuation}

An alternative way to extract $A(\omega)$ from $G(\tau)$ is the stochastic analytical continuation~\cite{arXiv:0403055}. Unlike the maximum entropy method, the stochastic analytical continuation does not depend on any \emph{a priori} parameters. It has been argued that the stochastic analytical continuation can produce more accurate spectral functions with more subtle structures. In the \texttt{HIBISCUS} component, we also implemented the stochastic analytical continuation which can be viewed as a useful complementary procedure to the maximum entropy method. Since the stochastic analytical continuation is computationally much heavier than the maximum entropy method, we parallelized it with MPI and OpenMP.

\underline{Kramers-Kronig transformation}

Once the analytical continuation is finished, we can obtain the spectral function $A(\omega)$ and the imaginary part of the real-frequency Green's function $\Im G(\omega)$,
\begin{equation}
A(\omega) = -\frac{\Im G(\omega)}{\pi}.
\end{equation}
From the well-known Kramers-Kronig transformation, the real part of $G(\omega)$ can be determined as well:
\begin{equation}
\Re G(\omega) = -\frac{1}{\pi} \int^{\infty}_{-\infty} \text{d}\omega^{\prime} \frac{\Im G(\omega)}{\omega - \omega^{\prime}}.
\end{equation}
In the \texttt{HIBISCUS} component, we offer a utility program to do this job.

\underline{Analytical continuation for the self-energy function: Pad\'{e} approximation}

To calculate real physical quantities, such as the optical conductivity, Seebeck coefficient, electrical resistivity, etc., the self-energy function on the real axis is an essential input. With the Pad\'{e} approximation~\cite{pade}, we can convert the self-energy function from the Matsubara frequency to the real frequency axis. We implemented the Pad\'{e} approximation for $\Sigma(i\omega_n)$ in the \texttt{HIBISCUS} component.

\underline{Analytical continuation for the self-energy function: Gaussian polynomial fitting}

The calculated results for the self-energy function on the real axis using the Pad\'{e} approximation strongly depend on the numerical accuracy of the input self-energy data. However, the CT-HYB/DMFT calculations usually yield a Matsubara self-energy function with significant noise~\cite{PhysRevB.76.205120}. In this case, the Pad\'{e} approximation does not work so well. To overcome this problem, K. Haule \emph{et al.}~\cite{PhysRevB.81.195107} suggested to split the Matsubara self-energy function into a low-frequency part and a high-frequency tail. The low-frequency part is fitted by some sort of model functions which depends on whether the system is metallic or insulating, and the high-frequency part is fitted by modified Gaussian polynomials. It was shown that their trick works quite well even when the original self-energy function is noisy, and is superior to the Pad\'{e} approximation in most cases. Thus, in the \texttt{HIBISCUS} component, we also implemented this algorithm. It has broad applications in the context of LDA + DMFT calculations~\cite{RevModPhys.78.865}.

\subsection{Application programming interface\label{subsec:api}}

We can not only execute the components of the $i$QIST software package directly, but also invoke them from external programs. To achieve this, we provide simple application programming interfaces (APIs) for most of the components in the $i$QIST software package for the Fortran and Python languages. With these well-defined and easy-to-use APIs, one can easily set up, start, and stop the CT-HYB impurity solvers. For example, one can use the following Python script fragment to start the CT-HYB impurity solver:
\begin{verbatim}
    from mpi4py import MPI           # import mpi support
    from pyiqist import api as ctqmc # import python api for iQIST
    ...
    comm = MPI.COMM_WORLD            # get the mpi communicator
    ctqmc.init_ctqmc(comm.rank, comm.size) # init. ctqmc impurity solver
    ctqmc.exec_ctqmc(1)              # exec. ctqmc impurity solver
    ctqmc.stop_ctqmc()               # stop  ctqmc impurity solver
\end{verbatim}
When the computations are finished, one can also collect and analyze the calculated results with Python scripts. Using these APIs, we have more freedom to design and implement very complex computational procedures. Please see Sec.~\ref{subsec:papi} for more details.

\section{Installation and usage\label{sec:install}}

In this section, we will explain how to install and use the $i$QIST software package.

\subsection{Get $i$QIST\label{subsec:get}}
The $i$QIST is an open source free software package. We release it under the GNU General Public Licence 3.0 (GPL). The readers who are interested in it can write a letter to the authors to request an electronic copy of the newest version of $i$QIST, or they can download it directly from the public code repository:
\begin{verbatim}
    http://bitbucket.org/huangli712/iqist.
\end{verbatim}

\subsection{Build $i$QIST\label{subsec:build}}
In order to build and install $i$QIST sucessfully, a Fortran 90 compiler (MPI-enabled), BLAS, and LAPACK linear algebra libraries are necessary. The components in $i$QIST can be successfully compiled using a recent Intel Fortran compiler. Most of the MPI implementations, such as MPICH, MVAPICH, OpenMPI and Intel MPI are compatible with $i$QIST. As for the BLAS implementation, we strongly recommend OpenBLAS. For the LAPACK, the Intel Math Kernel Library is undoubtedly a good candidate. Of course, it is also possible to use the linear algebra library provided by the operating system, for example, the vecLib Framework in the Mac OS X system. Some post-processing scripts contained in the \texttt{HIBISCUS} component are developed using Python. In order to execute these scripts or use the Python binding for $i$QIST, one should ensure that Python 2.x or 3.x is installed. Furthermore, the latest numpy, scipy, and f2py packages are also necessary.

The downloaded $i$QIST software package is likely a compressed file with zip or tar.gz suffix. One should uncompress it at first:
\begin{verbatim}
    $ tar xvfz iqist.tar.gz
\end{verbatim}
where \$ is the command line prompt. Then go to the iqist/src/build directory (in the following we just assume the top directory for $i$QIST software package is iqist) and edit the make.sys file to configure the compiling environment. One must set up the Fortran compiler, BLAS and LAPACK libraries manually:
\begin{verbatim}
    $ cd iqist/src/build
    $ editor make.sys
\end{verbatim}
Once the compiling environment is configured, please type the following command in the current directory (iqist/src/build) to compile $i$QIST:
\begin{verbatim}
    $ make all 
\end{verbatim}
After a few minutes (depending on the performance of compiling platform), if there are no error messages, all of the $i$QIST components are successfully compiled. 

Note that what you obtain are a few standalone applications. You can execute them in the terminal directly. If you want to compile them to a library, please edit the make.sys file again to active the API and MPY flags, and then re-compile the $i$QIST:
\begin{verbatim}
    $ editor make.sys
    $ make clean  (this step is optional)
    $ make lib
\end{verbatim}
At this time the libctqmc.a is generated. Then you can link it with your own Fortran programs. If you want to generate the Python binding for $i$QIST, please change the current directory to iqist/src/api:
\begin{verbatim}
    $ cd ../api
\end{verbatim}
and then use the following command to build pyiqist.so which is a valid Python module:
\begin{verbatim}
    $ make pyiqist
\end{verbatim}

\subsection{Setup $i$QIST\label{subsec:setup}}
Here we assume that the $i$QIST is built properly. Next we have to do one more step to finalize the installation. Please go to the iqist/bin directory and run the setup.sh:
\begin{verbatim}
    $ cd iqist/bin
    $ ./setup.sh
\end{verbatim}
If everything is OK, all of the executable programs, libraries, scripts, and Python modules will be collected and copied into the iqist/bin directory. Please add this directory into the system environment variables PATH and PYTHONPATH. Now the $i$QIST is ready for use.

\subsection{Use $i$QIST\label{subsec:usage}}
(i) At first, since there are several CT-HYB impurity solvers in the package and their features and efficiencies are somewhat different, it is the user's responsibility to choose suitable CT-HYB components to deal with the impurity problem at hand. (ii) Second, the $i$QIST is in essence a computational engine, so the users have to prepare scripts or programs to execute the selected CT-HYB impurity solver directly or to call it using the APIs. For example, if the users want to conduct CT-HYB/DMFT calculations, they must implement the DMFT self-consistent equation by themselves (The $i$QIST software package also provide a mini DMFT self-consistent engine for the Hubbard model on the Bethe lattice). (iii) Third, an important task is to prepare proper input data for the selected CT-HYB impurity solver. The optional input for the CT-HYB impurity solver is the hybridization function [$\Delta(i\omega_n)$], impurity level ($E_{\alpha\beta}$), interaction parameters ($U$, $J$, and $\mu$), etc. If the users do not feed these data to the impurity solver, it will use the default settings automatically. Specifically, if the Coulomb interaction matrix is general, one should use the \texttt{JASMINE} component to diagonalize the local atomic problem at first to generate the necessary eigenvalues and eigenvectors. (iv) Fourth, execute the CT-HYB impurity solver. (v) Finally, when the calculations are finished, one can use the tools contained in the \texttt{HIBISCUS} component to post-process the output data, such as the imaginary-time Green's function $G(\tau)$, Matsubara self-energy function $\Sigma(i\omega_n)$, and other physical observables. For more details, please refer to the user manual of $i$QIST.

\section{Examples\label{sec:examples}}
In the last few years, the $i$QIST software package has been successfully used in many projects, such as the study of the pressure-driven orbital-selective Mott metal-insulator transition in cubic CoO~\cite{PhysRevB.85.245110}, the metal-insulator transition in a three-band Hubbard model with or without SOC~\cite{PhysRevB.86.035150,du:2013}, the non-Fermi-liquid behavior in cubic phase BaRuO$_{3}$~\cite{PhysRevB.87.165139}, dynamical screening effects in the electronic structure of the strongly correlated metal SrVO$_{3}$ and local two-particle vertex functions~\cite{0295-5075-99-6-67003}, the electronic excitation spectra of the five-orbital Anderson impurity model~\cite{PhysRevB.89.245104}, an extended dynamical mean-field study of the 2D/3D Hubbard model with long range interactions~\cite{PhysRevB.90.195114}, electronic structures of the topological crystalline Kondo insulators YbB$_{6}$ and YbB$_{12}$~\cite{PhysRevLett.112.016403}, and superconducting instabilities of a multi-orbital system with strong SOC (doped Sr$_{2}$IrO$_{4}$)~\cite{Meng14b,PhysRevLett.113.177003}, etc. In order to illustrate the basic usage of the $i$QIST software package, we describe here several easily repeatable and simple applications of it. The testing platform is a Macbook laptop (CPU: Intel Core i7 2.3 GHz, Memory: 8 GB DDR3). We compile the $i$QIST software package using Intel Fortran Compiler 13.0.0 and the linear algebra library is Intel MKL. 

\subsection{Single-band Hubbard model\label{subsec:1band}}
\begin{figure}[ht]
\centering
\includegraphics[width=1.0\textwidth]{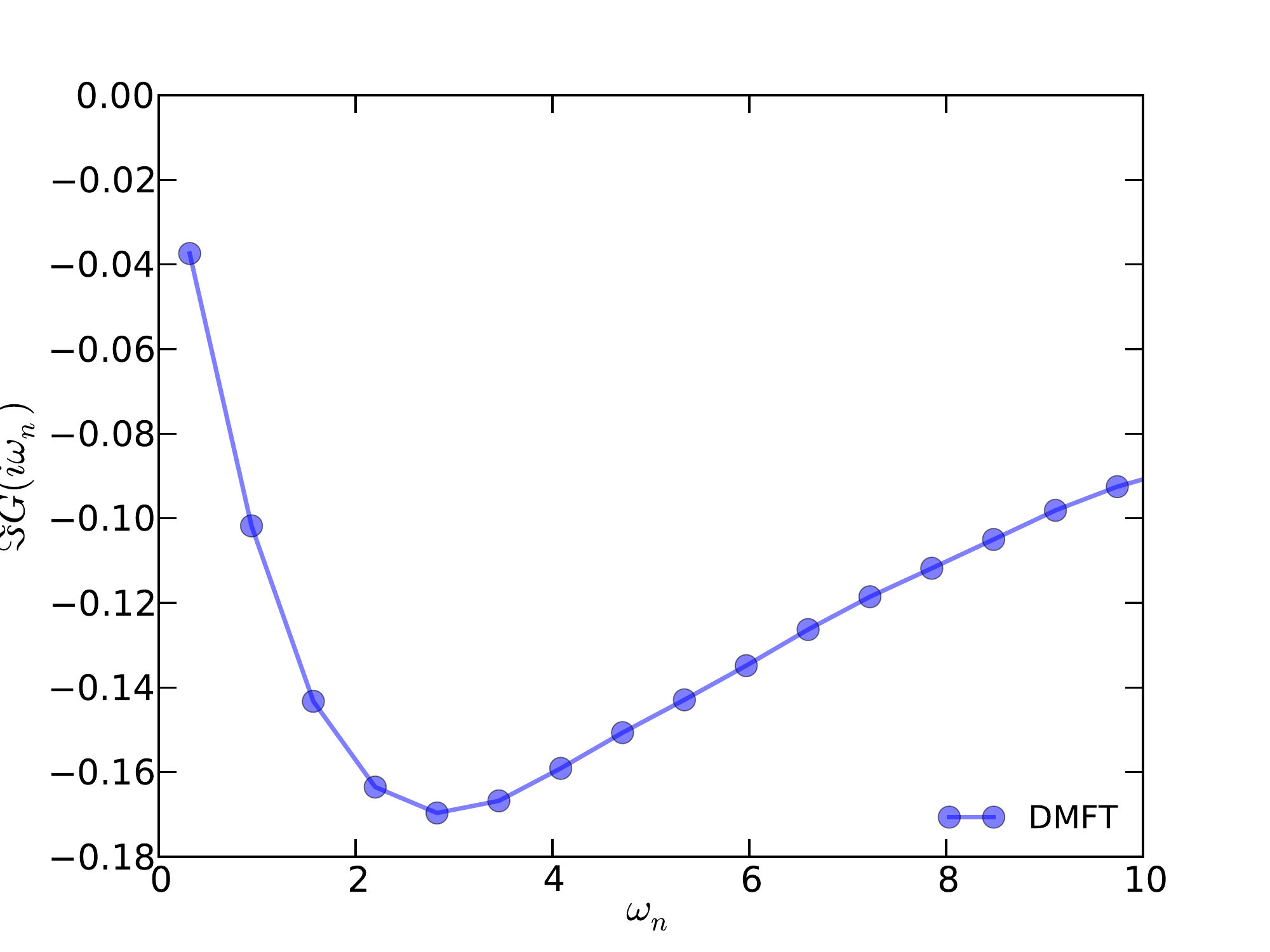}
\caption{Imaginary part of the impurity Green's function $\Im G(i\omega_n)$ of the single-band Hubbard model solved by DMFT. The model parameters are $U = 6.0$, $\mu = 3.0$, $\beta = 10.0$, $t = 0.5$. \label{fig:p1_g}}
\end{figure}
Here we consider the simplest case -- the single-band half-filled Hubbard model on the Bethe lattice. The model parameters are: Coulomb interaction $U = 6.0$, chemical potential $\mu = 3.0$, system temperature $T = 0.1$, hopping parameter $t = 0.5$. As mentioned before, we have implemented the DMFT self-consistency condition for the Bethe lattice ($\Delta = t^2 G$)~\cite{RevModPhys.68.13}, so we use $i$QIST to solve this model directly. The input file is as follows:
\begin{verbatim}
    # file name: solver.ctqmc.in 
    isscf = 2    ! control the running mode, self-consistent calculation
    isbin = 1    ! control the running mode, no data binning
    Uc    = 6.0  ! Coulomb interaction
    mune  = 3.0  ! chemical potential
    beta  = 10.0 ! inversion of temperature
\end{verbatim}
Note that the filename for the input file must be solver.ctqmc.in. Anything after the \# or ! character will be considered as comments and be skipped completely. Blank lines or even a blank solver.ctqmc.in file is valid. We choose the `key = value' or `key : value' format to set up the computational parameters. We do not need to set up all of the computational parameters in the solver.ctqmc.in file. They all have default values. As for the detailed explanations for the file format of solver.ctqmc.in and accurate definitions of all input parameters, please refer to the corresponding user manual encapsulated in the $i$QIST software package.

Now we choose the \texttt{AZALEA} component to solve this model. In order to reduce the numerical noise, 4 MPI processes are used:
\begin{verbatim}
    $ mpiexec -n 4 iqist/bin/azalea.x
\end{verbatim}
it takes about 2 minutes to complete this task. The calculated impurity Green's function (stored in the solver.grn.dat file), which exhibits clear insulating behavior, is shown in Fig.~\ref{fig:p1_g}. Finally, we should emphasize that the \texttt{GARDENIA} and \texttt{NARCISSUS} components are also applicable. The only thing we have to do is use gardenia.x or narcissus.x to replace azalea.x in the above command.
 
\subsection{Multiband Hubbard model with general Coulomb interaction}
\begin{figure}[ht]
\centering
\includegraphics[width=1.0\textwidth]{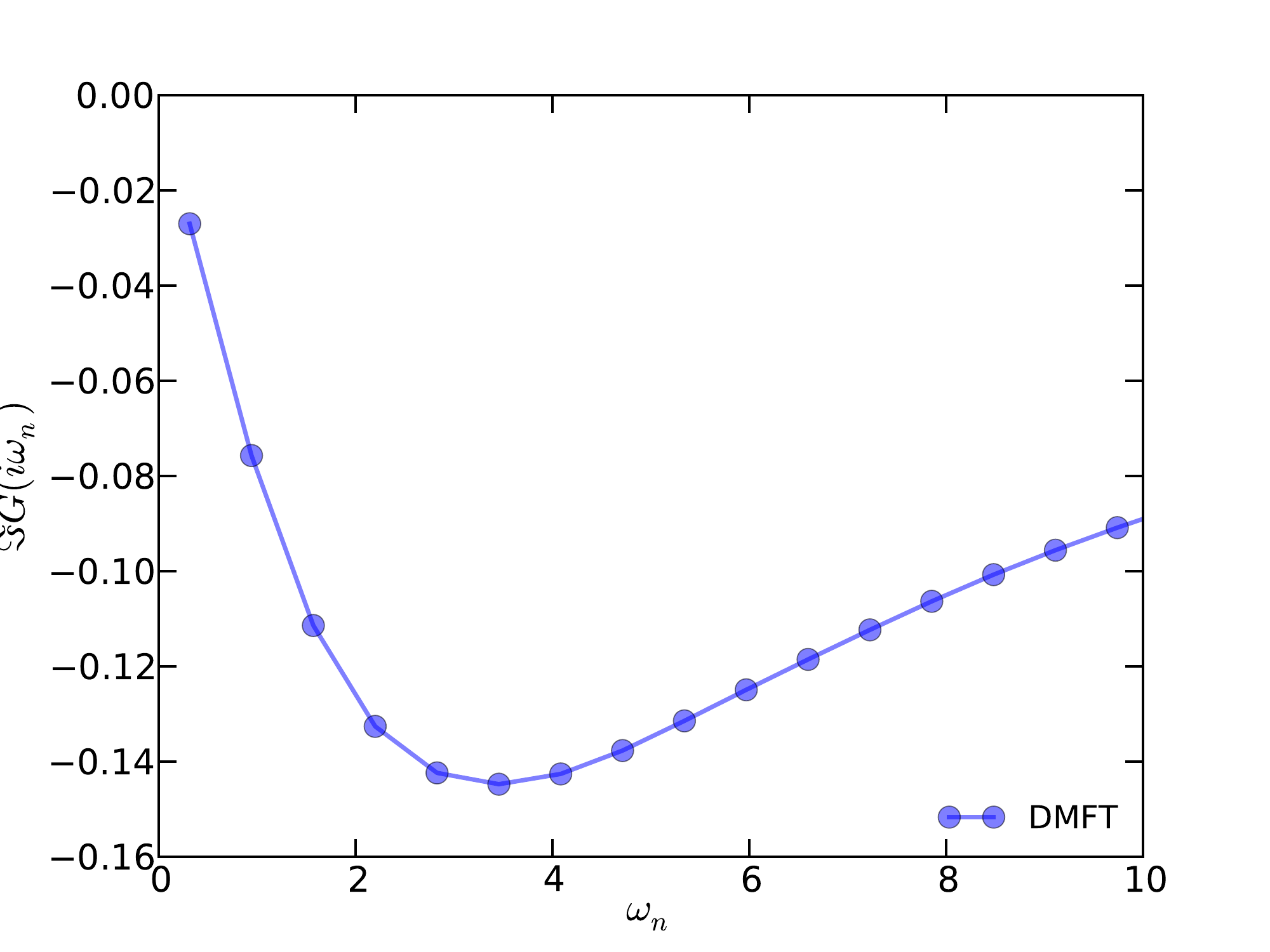}
\caption{Imaginary part of the impurity Green's function $\Im G(i\omega_n)$ of the two-band Hubbard model solved by DMFT. The model parameters are $U = 6.0$, $J_z = J_s = J_p = 1.0$, $\mu = 6.5$, $\beta = 10.0$, $t = 0.5$. \label{fig:p2_g}}
\end{figure}
Next we consider a two-band Hubbard model with rotationally invariant interaction on the Bethe lattice. The model parameters are: Coulomb interaction $U = 6.0$, Hund's exchange $J = 1.0$, chemical potential $\mu = 6.5$, system temperature $T = 0.1$, hopping parameter $t = 0.5$.

Since the interaction term is not of density-density type anymore, we have to use the general matrix version of the CT-HYB impurity solver, i.e., the \texttt{BEGONIA}, \texttt{LAVANDER}, \texttt{PANSY}, or \texttt{MANJUSHAKA} component to solve it. The atom.cix file which contains the eigenvalues and eigenvectors of local atomic problem, are nessary for these impurity solvers. So we have to generate the atom.cix file using the \texttt{JASMINE} component at first. The input file for the \texttt{JASMINE} must be atom.config.in. The required atom.config.in file is as follows:
\begin{verbatim}
    # file name: atom.config.in
    nband  : 2    # number of bands
    norbs  : 4    # number of orbitals (include spin index)
    ncfgs  : 16   # number of atomic configurations (= 2**norbs)
    nmini  : 0    # minmum occupancy
    nmaxi  : 4    # maximum occupancy
    Uc     : 6.00 # intraorbital Coulomb interaction
    Uv     : 4.00 # interorbital Coulomb interaction
    Jz     : 1.00 # z component of Hund's exchange interaction 
    Js     : 1.00 # spin-flip
    Jp     : 1.00 # pair-hopping
\end{verbatim}
We execute the \texttt{JASMINE} code in the command line:
\begin{verbatim}
    $ iqist/bin/jasmine.x   (the jasmine code is not parallelized)
\end{verbatim}
The key output files is atom.cix. Please do not modify it manually.

Here we select the \texttt{BEGOINA} component to solve this model. The corresponding input file looks as follows:
\begin{verbatim}
    # file name: solver.ctqmc.in
    isscf  : 2    ! control the running mode, self-consistent calculation
    isbin  : 1    ! control the running mode, no data binning
    nband  : 2    ! number of bands
    norbs  : 4    ! number of orbitals (include spin index)
    ncfgs  : 16   ! number of atomic configurations (= 2**norbs)
    mune   : 6.50 ! chemical potential for half-filling case
    beta   : 10.0 ! inversion of temperature
\end{verbatim}
You can see that in the solver.ctqmc.in file, the parameters for the Coulomb interaction and Hund's exchange interaction are absent. This is because the information about the local interaction has been included in the atom.cix file already.

Next let's conduct the calculation using MPI:
\begin{verbatim}
    $ mpiexec -n 4 iqist/bin/begonia.x
\end{verbatim}
The running time is about 16 minutes. In Fig.~\ref{fig:p2_g}, the obtained impurity Green's function is shown as a reference. In this example, we can use the \texttt{LAVENDER} component as well. With it we can adopt the orthogonal polynomial algorithm to improve the numerical accuracy and reduce the data noise.
 
\subsection{Two-particle Green's function and vertex function}
\begin{figure}[ht]
\centering
\includegraphics[width=0.55\textwidth]{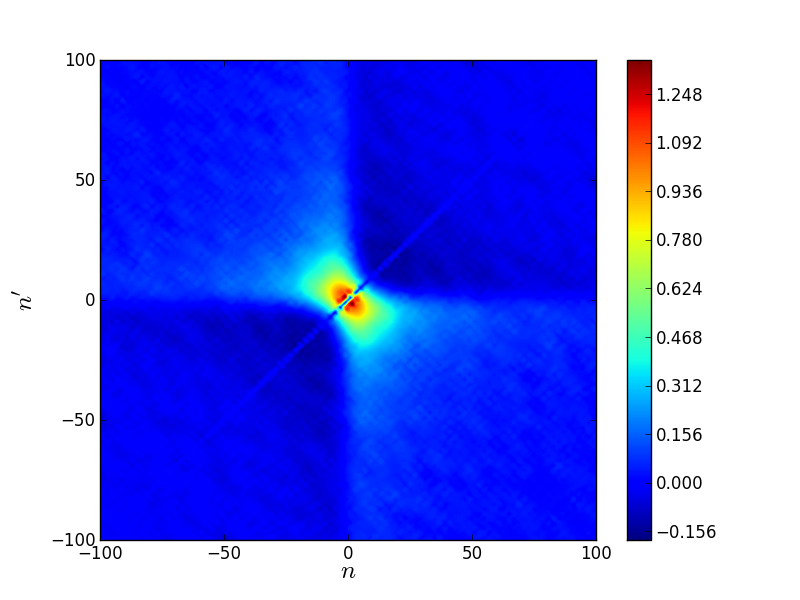}
\includegraphics[width=0.55\textwidth]{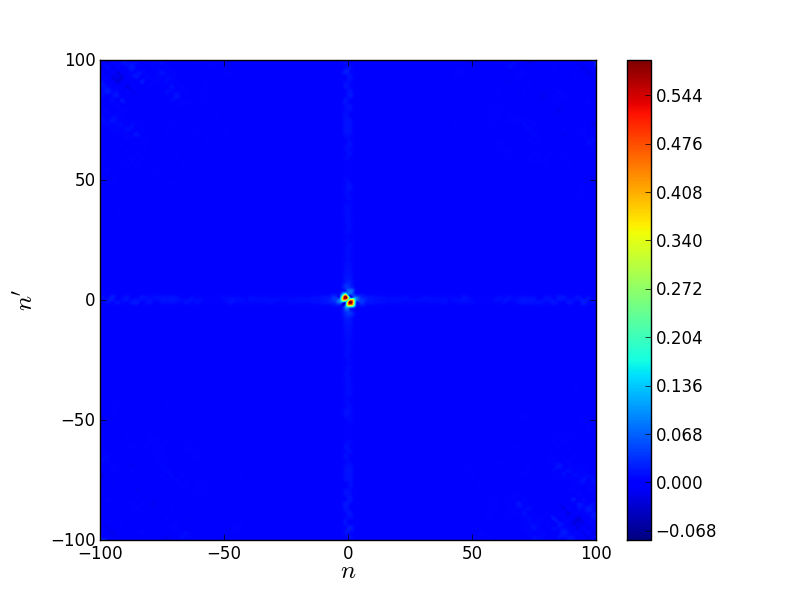}
\caption{Two-particle quantities of the single-band Hubbard model solved by DMFT. (Top) Two-particle Green's function $\Re \chi_{\uparrow\uparrow}(\omega_n, \omega_{n'}, \nu = 0)$. (Bottom) Two-particle vertex function $\Re \Gamma_{\uparrow\uparrow}(\omega_n, \omega_{n'}, \nu = 0)$. The model parameters are $U = 6.0$, $\mu = 3.0$, $\beta = 10.0$, $t = 0.5$. \label{fig:p3_cg}}
\end{figure}
In the previous two examples, DMFT self-consistent calculations are performed. Here we will show how to use $i$QIST to perform one-shot calculation to measure the two-particle Green's function and vertex function for a given impurity model.

For simplicity, we consider the same model as Sec.~\ref{subsec:1band} which was solved using the \texttt{AZALEA} component already. The converged hybridization function $\Delta(i\omega_n)$ is stored in the solver.hyb.dat file. Please copy it to the current directory and rename it to solver.hyb.in. Next, we prepare the solver.ctqmc.in file for the CT-HYB impurity solver:
\begin{verbatim}
    # file name: solver.ctqmc.in
    isscf = 1    # control the running mode, one-shot calculation
    isbin = 1    # control the running mode, no data binning
    isvrt = 8    # calculate two-particle quantities
    Uc    = 6.0  # Coulomb interaction
    mune  = 3.0  # chemical potential
    beta  = 10.0 # inversion of temperature
    nbfrq = 1    # number of bosonic frequencies
    nffrq = 128  # number of fermionic frequencies
\end{verbatim}
Since we are going to get the two-particle Green's function $\chi(\omega, \omega', \nu)$ and vertex function $\Gamma(\omega, \omega', \nu)$, the \texttt{GARDENIA} component is the best (the \texttt{NARCISSUS} component is OK, but it is less efficient than \texttt{GARDENIA}). We then use the following command to invoke it:
\begin{verbatim}
    $ mpiexec -n 4 iqist/bin/gardenia.x
\end{verbatim}
After about 10 minutes, the calculation is finished. The calculated two-particle quantities (stored in solver.twop.dat file) are shown in Fig.~\ref{fig:p3_cg} in which only the real part of the spin-up-up component is displayed.

\subsection{Python API\label{subsec:papi}}
\begin{figure}[tp]
\centering
\includegraphics[width=1.0\textwidth]{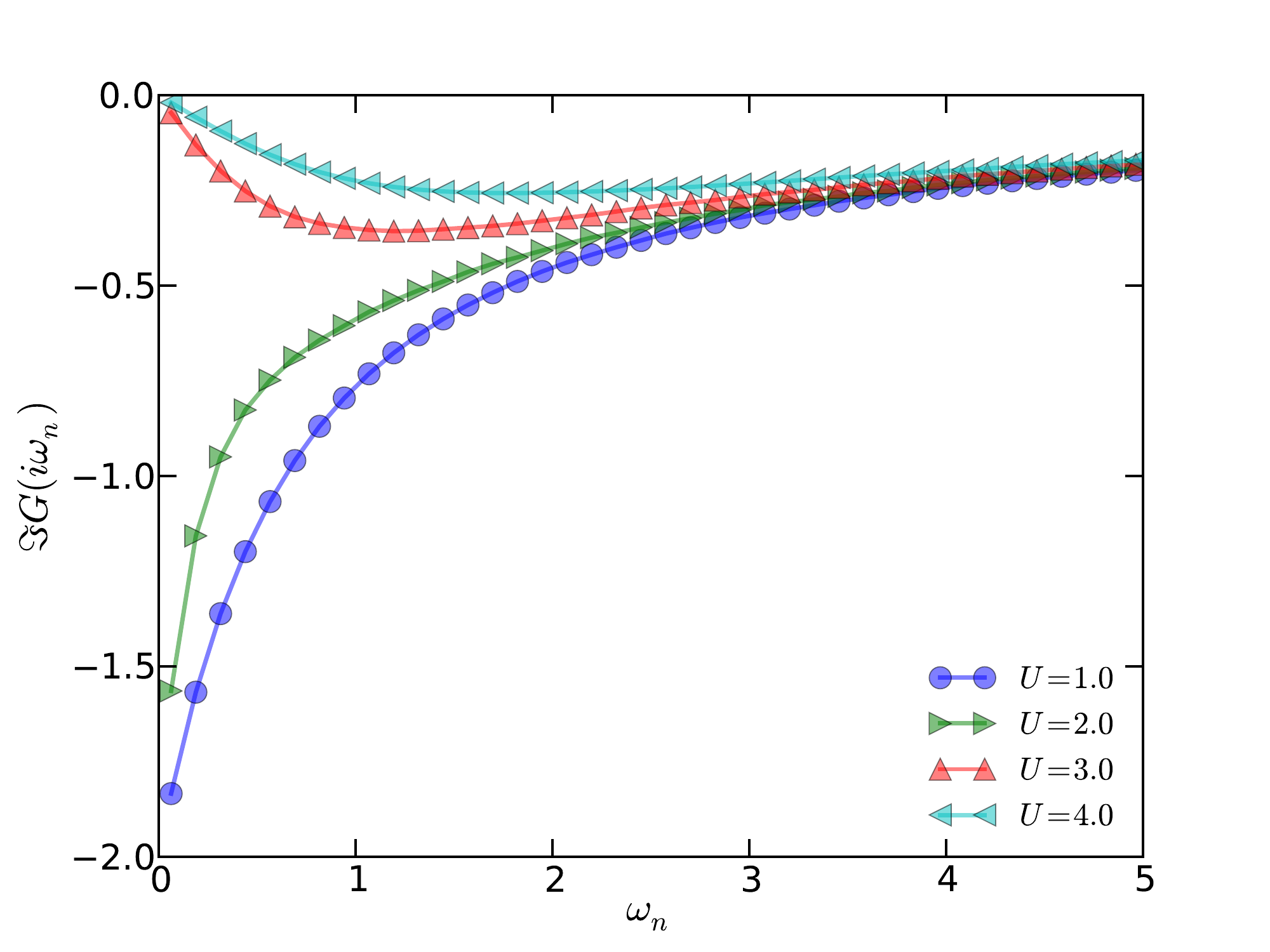}
\caption{Imaginary part of the impurity Green's function $\Im G(i\omega_n)$ of the single-band Hubbard model solved by DMFT. The model parameters are $\mu = U/2$, $\beta = 50.0$, $t = 0.5$. \label{fig:p4_g}}
\end{figure}
\begin{figure}[ht]
\centering
\includegraphics[width=1.0\textwidth]{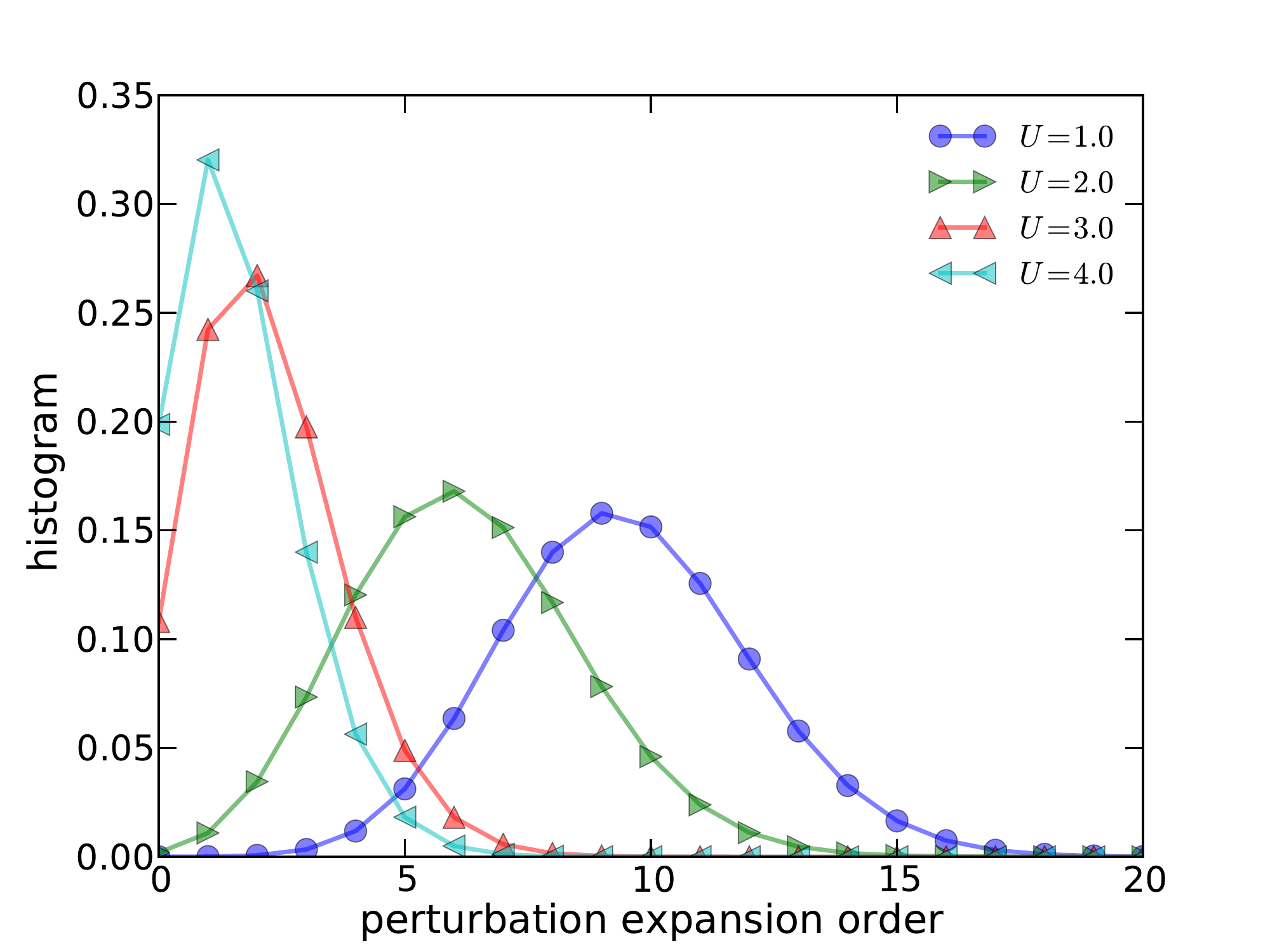}
\caption{Histogram of the perturbation expansion order for the single-band Hubbard model solved by DMFT. The model parameters are $\mu = U/2$, $\beta = 50.0$, $t = 0.5$. \label{fig:p4_h}}
\end{figure}
In the previous examples, we always execute the CT-HYB impurity solver components directly. However, $i$QIST provides flexible and powerful APIs for the Fortran and Python languages. We can use these APIs to develop complex computational programs easily. In this subsection, we try to use the Python binding of $i$QIST to build a somewhat complicated DMFT program, and use it to study the classic Mott-Hubbard metal-insulator transition in the single-band Hubbard model. The model parameters are $U = 1.0 \sim 4.0$, $\mu = U / 2 $, $\beta = 50.0$, $t = 0.5$.

Here is the full source code of the Python script:
\begin{verbatim}
    #!/usr/bin/env python
    import numpy           # import array support
    import shutil          # import high-level file operation support
    from mpi4py import MPI # import mpi support

    from u_ctqmc import *  # import the writer for solver.ctqmc.in file
    from pyiqist import api as ctqmc # import python module for iqist

    # get mpi communicator
    comm = MPI.COMM_WORLD

    # set up the basic parameters and allocate memory
    mfreq = 8193 # number of matsubara frequency points
    norbs = 2    # number of orbitals
    size_t = mfreq * norbs * norbs
    hybf_s = numpy.zeros(size_t, dtype=numpy.complex)

    # loop over Coulomb interaction strength: from 1.0 to 4.0
    for u in range(1,5):
        # build ctqmc input file: solver.ctqmc.in
        if comm.rank == 0: # only the master process can do it
            p = p_ctqmc_solver('azalea') # select impurity solver
            p.setp(isscf = 1, isbin = 1) # set up parameters
            p.setp(beta = 50.0)          # set up parameters
            p.setp(Uc = u, mune = u/2.0) # set up parameters
            p.write()                    # write solver.ctqmc.in
            del p
        comm.Barrier() # mpi barrier

        # DMFT self-consistent loop
        ctqmc.init_ctqmc(comm.rank, comm.size) # init ctqmc impurity solver
        for i in range(20): # number of iterations = 20
            ctqmc.exec_ctqmc(i+1)         # execute ctqmc impurity solver
            grnf = ctqmc.get_grnf(size_t) # get impurity Green's function
            hybf = (0.25*grnf+hybf_s)/2.0 # DMFT self-consistent condition
            hybf_s = hybf                 # update old hybridization function
            ctqmc.set_hybf(size_t, hybf)  # set up hybridization function
        ctqmc.stop_ctqmc() # stop ctqmc impurity solver
        comm.Barrier() # mpi barrier

        # save calculated results
        if comm.rank == 0: # only the master process can do it
            shutil.move('solver.grn.dat','solver.grn.dat.'+str(u))
            shutil.move('solver.hist.dat','solver.hist.dat.'+str(u))
\end{verbatim}
In this Python script (dmft.py), the pyiqist module contains the Python binding for $i$QIST which is introduced in Sec.~\ref{subsec:api}. The u\_ctqmc module which implements the p\_ctqmc\_solver class is included in the \texttt{HIBISCUS} component and is often used to generate solver.ctqmc.in file dynamically. The MPI parallelism is fully supported in this script via the mpi4py module. To run it, please use the following command:
\begin{verbatim}
    $ mpiexec -n 4 ./dmft.py
\end{verbatim}
It takes about half an hour to finish this job. The calculated results (the solver.grn.dat file contains the impurity Green's function, and the solver.hist.dat file contains the histogram data) are shown in Fig.~\ref{fig:p4_g} and Fig.~\ref{fig:p4_h}, respectively. Clearly, between $U = 2.0$ and $U = 3.0$, a Mott metal-insulator transition induced by electronic interaction occurs. And the perturbation expansion order of CT-HYB impurity solver decreases with the increment of interaction strength.

\section{Future developments\label{sec:conclusion}}
In this paper, we explained and demonstrated the $i$QIST software package. $i$QIST aims to provide a complete toolkit for solving various quantum impurity systems. At first, we introduced the basic theory about quantum impurity models and the CT-QMC/CT-HYB algorithm briefly. And then various optimization tricks and algorithms implemented in $i$QIST have been discussed in detail. Following that we reviewed the software architecture and major features of $i$QIST. The compiling, setup, and workflow of $i$QIST were also illustrated. Finally, several simple examples have been shown to help the readers master the basic usage of $i$QIST step by step. 

Although proven to be very versatile in applications and efficient in performance, the $i$QIST project is still a work in progress and the development will continue. The future developments of the $i$QIST project are likely to be along the following directions.

As the study of interacting electronic systems is moving towards treating their correlated multi-band nature in a more realistic fashion (5- or 7-bands, SOC included, competing multi-orbital interactions, etc.), it is important to develop even more efficient and optimized CT-HYB impurity solvers. An effective way to reduce the average size of the matrices used during the calculation is to fully consider the point group symmetry of the impurity model, which provides more GQNs to the problem. The corresponding coding work has already been started by some of the authors.

Recent developments in condensed matter theories need to be added into the features of the $i$QIST software package. For example, the measurement of entanglement entropy in realistic correlated fermion systems~\cite{PhysRevLett.111.130402,PhysRevB.89.125121,PhysRevLett.113.110401} will be considered, with which one will be able to explore and discovery more symmetry protected topological states and even interaction-driven topological orders that might exist in nature~\cite{Chen21122012,Wang07022014}. 

The two-particle correlation functions (susceptibilities) contain more information than the single-particle quantities, but the DMFT formalism is only self-consistent at the single-particle level. To conduct a calculation which is self-consistent both at the single- and two-particle levels is the next step in the CT-HYB/DMFT simulations. The DMFT + Parquet scheme present in Ref.~\cite{Meng14b} and \cite{PhysRevLett.113.177003} is the first step to incorporate correlation effects at the two-particle level beyond single-site DMFT, but it is only self-consistent at the two-particle level, and in many occasions only one-shot simulations at the two-particle level are considered due to numerical difficulties. To be fully self-consistent among single- and two-particle quantities, one still needs to employ the Schwinger-Dyson equation to feed the two-particle information back to the single-particle quantities~\cite{PhysRevE.80.046706,PhysRevE.87.013311}. This will also be a further development of the $i$QIST software package.

Instead of using single- and two-particle diagrammatic relations to capture the spatial correlation effects, one can also develop cluster CT-QMC impurity solvers, such that the spatial correlations within the cluster can be captured exactly. While in one-band models and a few two-band models cluster CT-QMC impurity solvers are available~\cite{RevModPhys.77.1027,RevModPhys.78.865,PhysRevB.88.041103,PhysRevB.88.245110,PhysRevB.89.195146}, generic cluster CT-QMC impurity solvers which take care of both the multi-orbital interactions within each cluster site and the spatial correlations between the cluster sites are still missing. This is also an arena for future developments.

In the end, we would like to emphasize that $i$QIST is an open initiative and the feedback and contributions from the community are very welcome.

\section*{Acknowledgments}
YLW, LD and XD are supported by the National Science Foundation of China and the 973 program of China (No. 2011CBA00108). Their calculations were preformed on TianHe-1A, the National Supercomputer Center in Tianjin, China. ZYM thanks the inspiring guidance from H.-Y. Kee and Y. B. Kim for bringing his attention to multi-orbital physics, he acknowledges the NSERC, CIFAR, and Centre for Quantum Materials at the University of Toronto, and the National Thousand-Young-Talents Program of China. His computations were performed on the GPC supercomputer at the SciNet HPC Consortium. LD acknowledge financial support through DARPA Grant No. D13AP00052. LH and PW acknowledge support from the Swiss National Science Foundation (Grant No. 200021\_140648).

\clearpage

\bibliographystyle{apsrev4-1}
\bibliography{iqist}

\end{document}